\documentclass[sigconf]{acmart}

\AtBeginDocument{%
  \providecommand\BibTeX{{%
    \normalfont B\kern-0.5em{\scshape i\kern-0.25em b}\kern-0.8em\TeX}}}

\usepackage{soul}
\usepackage{color}
	\definecolor{celadon}{rgb}{0.67, 0.88, 0.69}
    \definecolor{flamingopink}{rgb}{0.99, 0.56, 0.67}

\usepackage{pbox}
\usepackage{makecell}
\usepackage{appendix}

\usepackage{footnote}
\usepackage{tablefootnote}
\makesavenoteenv{tabular}
\makesavenoteenv{table}

\copyrightyear{2020}
\acmYear{2020}
\setcopyright{acmlicensed}
\acmConference[ARES '20]{The 15th International ARES Conference on Availability, Reliability and Security}{August 25--28, 2020}{Dublin, Ireland}
\acmBooktitle{The 15th International ARES Conference on Availability, Reliability and Security, June 25--28, 2030, Dublin, Ireland}
\acmDOI{https://doi.org/10.1145/3407023.3407068}
\acmISBN{978-1-4503-8833-7/20/08}

\begin{document}

\title{SoK: Exploring the State of the Art and the Future Potential of Artificial Intelligence in Digital Forensic Investigation}



\author{Xiaoyu Du}
\affiliation{%
    \institution{University College Dublin}
    \country{Ireland}}
    \email{xiaoyu.du@ucdconnect.ie}

\author{Chris Hargreaves}
\affiliation{%
    \institution{University of Oxford}
    \country{United Kingdom}}
    \email{christopher.hargreaves@cs.ox.ac.uk}
    
\author{John Sheppard}
\affiliation{%
    \institution{Waterford Institute of Technology}
    \country{Ireland}}
    \email{jsheppard@wit.ie}
    
\author{Felix Anda}
\affiliation{%
    \institution{University College Dublin}
    \country{Ireland}}
    \email{felix.anda@ucdconnect.ie}
    
\author{Asanka Sayakkara}
\affiliation{%
    \institution{University College Dublin}
    \country{Ireland}}
    \email{asanka.sayakkara@ucdconnect.ie}
    
\author{Nhien-An Le-Khac}
\affiliation{%
    \institution{University College Dublin}
    \country{Ireland}}
    \email{an.lekhac@ucd.ie}
    
\author{Mark Scanlon}
\affiliation{%
    \institution{University College Dublin}
    \country{Ireland}}
    \email{mark.scanlon@ucd.ie}

%

\renewcommand{\shortauthors}{Du, Hargreaves, Sheppard, Anda, Sayakkara, Le-Khac and Scanlon}

\begin{abstract}

Multi-year digital forensic backlogs have become commonplace in law enforcement agencies throughout the globe. Digital forensic investigators are overloaded with the volume of cases requiring their expertise compounded by the volume of data to be processed. Artificial intelligence is often seen as the solution to many big data problems. This paper summarises existing artificial intelligence based tools and approaches in digital forensics. Automated evidence processing leveraging artificial intelligence based techniques shows great promise in expediting the digital forensic analysis process while increasing case processing capacities. For each application of artificial intelligence highlighted, a number of current challenges and future potential impact is discussed.

\end{abstract}

\keywords{Digital Forensics, Machine Learning, Deep Learning}


\maketitle

\section{Introduction}

Digital Forensic Science involves the recovery of evidence from digital devices, and is sometimes defined in terms of process models that capture the stages of the investigation~\cite{du2017processmodelsdfaas}. For the purposes of this paper, the process is divided up into stages that assist in the discussion of where AI techniques have been applied to a digital investigation. These are: acquisition, examination, analysis, and presentation, similar to the Interpol guidelines~\cite{Europol01}. This overall digital investigation process can be applied to a variety of data sources: traditional computers, mobile and other embedded devices (such as UAVs, smart home devices and other IoT devices). It can also apply to network forensics, cloud forensics, and live forensics.





This paper is structured into three main parts. Section 2 provides a brief introduction to artificial intelligence (AI) techniques, including the most useful references for a digital forensics researcher to become familiar with the area. Section 3 provides a series of sections, each describing a sub-area of the digital forensics field where AI techniques have been already applied. Each of these subsections has a consistent structure which starts with an introduction to the subtopic, a overview of the current AI applications in that area, and finishes with current challenges and future directions. Finally, Section 4 provides a general discussion of challenges and future directions for AI applications in digital forensics.

The contribution of this paper is therefore a comprehensive systematisation of AI research in digital forensics that can be used by digital forensic researchers and practitioners to identify the latest applications of AI in particular sub-areas, but also an important resource for AI researchers looking for real-world application areas for their new techniques and the challenges that are unique to applying AI in the specific field of digital forensics.








\section{Background on Artificial Intelligence}
AI, or machine intelligence, is the discipline studying intelligent agents, i.e., an agent that reacts to its environment to achieve an optimal path to its goal. In Computer Science, AI can be split into two primary fields; Machine Learning (ML) and Deep Learning (DL). The success of AI is data-driven in so far as no explicit code controls the precise output. The datasets used for training the models are critical, and data pre-processing is a key step in ML. An overview of the datasets available for training AI models in Digital Forensics is provided by~\citet{grajeda2017availability}. 

\subsection{Machine Learning}




ML has been widely applied to digital forensic investigation for data discovery~\cite{chen2018file, vulinovic2019neural}, device triage~\cite{marturana2011quantitative, marturana2013machine}, network forensics~\cite{Moustafa:2019}, etc. Flach~\cite{flach2012machine} outlined the ML ingredients as: \textbf{tasks}, the problems that can be solved; \textbf{models}, the output of ML; and \textbf{features}, the workhorses of ML. There are three steps for ML applications: 1) task definition; 2) feature construction; 3) evaluation and optimisation.

An ML task is an abstract representation of the problem. For a prediction problem, it can be defined to be either a classification/clustering or regression problem, depending on the type of target labels. Take age estimation as an example. If age is considered categorical, it can be defined as a classification task; while it could be a regression task if the age is numeric.

Feature construction is crucial for the success of ML application~\cite{flach2012machine}. There are different kinds of features: categorical, ordinal and quantitative. For text analysis, the raw data is a sequence of symbols cannot be fed directly to algorithms, \textit{bag-of-word} representation is applied. For image data, patch or contiguous patches can be extracted. During the experiment, features are transformed and selected to reducing over-fitting, improve performance or reduce training time. The \textit{No-Free-Lunch} theorem implies that there is no ultimate feature learner; it is variable depending on the data distribution and learning algorithm~\cite{shalev2014understanding}.

Models are the output of ML~\cite{flach2012machine}. Model evaluation enables its refinement, and the process is iterated until the performance is sufficient. A confusion matrix is able to show the accuracy of a classification task, where the classification performance of each class can be found. The F1 score is an average accuracy of each class, which shows the average performance of the model. Precision and recall are usually used in the evaluation matrix.




\subsection{Overview of Deep Learning}

The key differentiator of DL from ML is that the features are not designed by human engineers. Instead, they are learned from data using a general-purpose learning procedure~\cite{lecun2015deep}. ML tasks require input that is computationally convenient to process. However, it is often difficult to engineer features of real-world data such as images, video, and sensor data. Representation (feature) learning techniques employed by artificial neural networks (ANNs) allows a system to automatically discover the representations needed for feature detection or classification from raw data~\cite{lecun2015deep}.

A DL model can be described in two stages; optimisation and inference. The optimisation process, known as training, is used to update the weights connecting the layers of neurons defined in the model. The process of weight update is achieved by a back-propagation algorithm~\cite{lecun2015deep}. Before training a DL model, a loss objective is defined to measure the difference/error between the predicted outputs and the targets. The model updates its weights with the objective of minimising the loss function through many times of iterations. To make it closer to the objective, the mathematics under the hood are gradient descent algorithms for minimising the loss~\cite{ruder2016overview}. After completing the optimisation, then the model is applied for inference, namely, making predictions on data that are unseen during training. One key performance metric e is the generalisation ability. That says if the model generalises well, it performs on the unseen (test) data as well as the training data.




DL is often applied for natural language processing (NLP) and computer vision (CV), but more specific applications include content filtering~\cite{wang2014improving}, e-commerce recommendations~\cite{shankar2017deep}, and search result relevancy scoring~\cite{yin2016ranking}. Other applications that are discussed later include camera sensor model identification, image forgery detection, facial detection and recognition, text clustering, etc. Digital forensic specific applications include malware classification, network intrusion detection, file fragment typing, watermarking, steganalysis, pattern recognition, timeline analysis, etc.


\section{Applications of AI in DF}

\subsection{Data Discovery and Recovery}

One of the early stages of a digital investigation is making the digital evidence obtained available in a human readable form~\cite{kohn2013integrated} (extraction). This can include extracting information from known file systems and file types, but also recovering deleted data.


\subsubsection{State of the Art of AI in Data Discovery}

Files deleted within a file system may be recoverable deterministicly if some metadata remains. However, in some cases this metadata is absent and the file content resides in the unallocated parts of a volume. File carving is the process of recovering such files without the metadata. However, it is also possible that such files may be fragmented over the disk and partially overwritten. \citet{garfinkel2007carving} reported on fragmentation statistics collected from over 350 disks containing FAT, NTFS and UFS file systems. While fragmentation on a typical disk is low, the fragmentation rate of forensically important files such as email, JPEG and Word documents is relatively high.

As the search space for fragments belonging to a particular file is so large, distinguishing the file type of a fragment can shorten the search time. One approach proposed for file fragment classification used NLP~\cite{fitzgerald2012using}. In this research, a supervised learning approach is taken based on the use of support vector machines (SVM) combined with the \textit{bag-of-words} model. File fragments are represented as ``bags of bytes'' with feature vectors consisting of unigram and bigram counts as well as other statistical measurements (including entropy). \citet{chen2018file} proposed a novel scheme based on fragment-to-grayscale image conversion and DL to extract hidden features and therefore improve the accuracy of classification. This CNN model was trained and tested on the public \textit{GovDocs} dataset. The average classification accuracy achieved was 70.9\%. Vulinovic et al.~\cite{vulinovic2019neural}. \citet{vulinovic2019neural} applied a CNN model using 1D convolution on the original byte block. Both feedforward neural networks (FFNN) and CNNs are tested. FFNNs achieved better results using selected bigrams as input the highest macro-average F1 score being 0.8138. 

Another problem faced during file carving is to determine the ownership of carved information when the storage media is used by more than one user. An automated solution to the multi-user carved data ascription was proposed by~\citet{garfinkel2010automated}. The features used by the automated ascription system are 1) file system metadata (MAC timestamp, file owner), 2) file placement (i.e., sector, fragment) information , 3) embedded file metadata (JPEG camera model, Word file save time, etc.). The data used to verify this system is disk images from the Real Data Corpus ~\cite{garfinkel2009bringing}, a collection of more than 2,000 disk images made from hard drives that were purchased on the secondary market. The result shows accuracy of classification is from 65.66\% to 99.83\%. In the end, this approach achieved a low accuracy (0\%) considering no discernible difference between the activity patterns of each user. 

\subsubsection{Current Challenges and Future Directions}
The current literature shows automation in digital forensic investigation employing statistical measurement for data representation and ML algorithms for classification. ML techniques have the potential to acquire useful information for investigations more efficiently -- leveraging the accumulation of experience learned from the previous digital evidence analysis. Adversarial attacks are one of the challenges of AI model development. It has been suggested that the existence of adversarial attacks may be an inherent weakness of DL models~\cite{madry2017towards}. The adversary can manipulate the input resulting in incorrect output. Adversarial attacks could also be used as a counter forensics technique. As a result, any pre-trained model could loose its effectiveness during an investigation. To this end, anti-counter-forensics for adversarial attacks remains an open question.






  





\subsection{Device Triage}
With the proliferation of digital evidence, the data volumes encountered in investigations is a significant challenge faced by Law Enforcement Agencies (LEAs). Digital evidence triage was proposed for the timely identification, analysis, and interpretation of digital evidence, with a process model proposed in \citet{rogers2006computer}. Currently, the prioritisation of device acquisition and processing at a crime scene is determined by the investigative officer. As more AI based techniques are developed, on-scene preliminary inspections could quickly focus the analysis towards the devices most likely to contain case-progressing information first.




\subsubsection{State of the Art of AI in Device Triage}

With the increasing significance of mobile device forensics, \citet{marturana2011quantitative} proposed an approach for device prioritisation leveraging data mining and ML theory. This work presents the result of a study concerning mobile phone classification in a real child abuse investigation case. The features used consisted of the phone model, phone contacts, calls made, text messages sent/received/read, number of video/audio/photo files, URL, email, memos. The experimentation tested the performance on the feature value represented as numeric (a number) and category (the number is low, medium or high). 
 
In some subsequent work, \citet{marturana2013machine} expanded the triage approach to detect the device's relative importance using features from: \textit{1) the timeline of events, 2) the crime’s specific features, and 3) the suspect’s private sphere (habits, skills and interests)}. The experimentation in this work was conducted on a copyright infringement and a CSEM exchange case. The dataset applied consisted of 23 cell phones for the CSEM case with 13 digital media files and 45 copyright infringement-related features. A result of 99\% correctly classified samples on both cases was achieved.

\subsubsection{Current Challenges and Future Directions}

The lack of a sufficiently large, shared dataset is a challenge for developing AI triage models. As the triage task consists of a quick, simple examination and analysis to help investigators to reduce the noise and identify relevant information quickly, the development of a emulated, realistic dataset is a substantial task. 

Future digital investigation may heavily rely on efficient device triage. The report of serious digital forensics backlogs~\cite{scanlon2016battling} indicates comprehensive examination of all digital devices is almost impossible. Increasing the accuracy of triage would result in less resources wasted processing non-pertinent data. In addition, and common for all ML approaches, the training dataset determines the performance of the model. The higher the volume and quality of data used to train the model, the better the model will perform. 

Investigations involving multiple devices is common, if not the norm. Multiple device analysis and triage can be integrated. For example, when determining the importance of the devices, the actions between them can be considered, e.g., file sharing, device connections, information exchanging, etc.). 

\subsection{Network Traffic Analysis}


The voluminous nature of data associated with Network Traffic Analysis (NTA) makes it an excellent candidate for the application of AI techniques to help filter redundant information and automate the detection of crimes or other forms of misconduct. 

\subsubsection{State of the Art of AI in Network Traffic Analysis}

Network investigations often form a part of a bigger investigation involving incident response, cloud, IoT, mobile devices, wearable technologies and fraudulent monetary activities. These investigations tend to involve multiple devices or technologies which have been communicating with each other. A wealth of literature is available to investigators for the use of Intrusion Detection techniques to network data offline in batch mode after the fact. Surveys on the use of AI in IDS can be found in~\citep{Moustafa:2019}, ~\citep{Othman:2018},~\citep{Buczak:2016} and~\citep{Ahmed:2016}. 

Feature selection techniques impact heavily on the models produced by AI techniques. The most up to date datasets for intrusion detection include the CICIDS 2017, CICIDS 2018~\citep{Sharafaldin:2018} and CICDDoS2019~\citep{Sharafaldin:2019} whose features are constructed using CICFlowMeter-V3~\citep{Lashkari:2017}. Principal Component Analysis (PCA) has been applied to these datasets. AI techniques used to model the CICIDS datasets include SVMs and DL~\citep{Aksu:2018b,Ustebay:2018,Vinayakumar:2019}. Auto-Encoders and PCA were used for dimensionality reduction in ~\cite{Abdulhammed:2019}. The reduced datasets were evaluated using classifiers such as Random Forest (RF), Bayesian Network (BN), Linear Discriminant Analysis (LDA) and Quadratic Discriminant Analysis (QDA). AI techniques have successfully been employed for Botnet Detection using ML on DNS requests~\citep{Biradar:2020,SINGH:2019}, while~\citep{ALAUTHMAN:2000} used traffic reduction with Reinforcement Learning (RL).

\citet{Elrawy:2018} surveyed the challenges of security in IoT and presented a comprehensive review of current anomaly-based IoT IDSs. \citet{Deng:2018} proposed a lightweight ML NIDS for IoT environments using a combination of fuzzy c-means clustering (FCM) and PCA, while~\citet{Amouri:2018} presented a NIDS with low computational and resource requirements using decision trees. A data mining approach for an IoT NIDS using PCA and suppressed fuzzy clustering (SFC) techniques proved to be well suited to high dimensional spaces producing high levels of accuracy~\cite{Liu:2018}. \citet{SAFAEIPOUR:2019} applied PCA for feature selection, with clustering techniques, to IoT data to infer exploited IoT devices, and IoT coordinated probing campaigns.

Indicators of Compromise (IOC) are evidence artefacts that are indicative of a system or network being attacked. Useful sources of this data can be network packets or network logs. In~\citep{Rene:2017}, network data was collected and features extracted before applying clustering techniques to extract IOC rules for malware detection.

Android network traffic was modelled for malware detection in~\cite{Murtaz:2018}. RF, K-Nearest Neighbour (KNN), decision Tree (DT), Random Tree (RT) and Regression were all applied to the CICAndMal2017 dataset which was generated and made available by the authors. An updated version of the dataset was evaluated by~\citet{Taheri:2019} using Random Forest classification. This also utilised API call data for classification. A DT approach for the detection of cryptocurrency miners is presented in~\cite{FITPUB12057}. 

\subsubsection{Current Challenges and Future Directions.}
Network traffic analysis is becoming increasingly hierarchical -- providing better potential for correlation of user data over multiple networks or devices. Modern networks and devices allow for the broader profiling of individual suspect users and their actions. Correlation of incidents in new and emerging environments should also be inter-device dependent. This is of particular importance in areas such as in the event of an modern automobile crash. Network traffic analysis will also see growth in inter-user correlation, e.g., the correlation of mobile phone communication through applications over data networks to identify who a suspect is in contact with most frequently or most recently relative to a certain time period. 


One of the biggest challenges network analysis faces is the huge increase in volumes of data from new and emerging devices that needs to be gathered, stored and modelled. Classification and prediction models require accurate and up to date datasets with the correct features. Datasets traditionally used in this area have suffered from problems such as a lack of relevant or real-world data, bias and disproportionate classes. In the area of IDS, these datasets by their nature are always behind the curve in terms of up to date attacks. This creates issues in the creation and evaluation of accurate models. GDPR legislation also raises issues around user privacy for inter-event correlation. Novel protocols associated with emerging devices can result in previously undocumented network traffic patterns. This can affect the performance of existing pre-trained AI models, which obviously may not have taken this new activity into account during training. Encryption poses a challenge to network traffic analysis but does not hinder it completely. Even with encrypted networks, AI techniques can still be used to model statistical information of a network. 


\subsection{Forensics on Encrypted Data}


One of the most significant issues facing digital forensics investigators around the globe is encrypted data. 
The prevalence of cryptographically protected devices and data poses an inevitable threat to digital forensic investigation. If the device under investigation uses disk encryption, the forensic disk image becomes unusable~\cite{lillis2016challenges}. Currently, the law of many countries demands that the owner of the device has to surrender their passwords/keys to LEAs under warrant. However, unavailability/non-compliance often brings the investigation of the encrypted device to a halt~\cite{vincze2016challenges}. Due to the large bit length used in modern cryptographic algorithms a successful brute-force attack is computationally infeasible.


Side-channel attacks on encryption algorithms have been proven as effective key attack vectors~\cite{spreitzer2018systematic}. An electromagnetic side-channel analysis (EM-SCA) attack is performed by observing the EM emissions over time of a device under test (DUT) while it is performing data encryption/decryption. A single such observation is called an \emph{EM trace} containing the three signal characteristics; \textit{Amplitude}, \textit{Phase}, and \textit{Frequency}. Once a sufficient number of EM traces are collected, they are fed into an EM-SCA algorithm, e.g., differential electromagnetic analysis (DEMA) or correlation electromagnetic analysis (CEMA), to extract the underlying cryptographic key~\cite{kocher1999differential, kocher2011introduction}. These algorithms require the EM traces to be precisely aligned in the time-domain in order to succeed. Due to the nature of EM trace extraction, minor misalignments often force the attacker to extract more EM traces -- this alignment issue can be greatly improved by ML~\cite{sayakkara2019survey}. Furthermore, these algorithms can take a considerable time to complete, making them difficult to be used in live device investigation scenarios in digital forensics~\cite{le2007efficient, tian2012clock, zhou2019deep}.

\subsubsection{State of the Art of AI on Handling Encrypted Data}

There are two potential avenues for EM-SCA that can be assisted by AI techniques; gaining useful insights without accessing the encrypted content, and performing cryptographic key retrieval attacks. Towards the first goal, various AI approaches have been applied using power and electromagnetic side-channel observational data~\cite{sayakkara2019survey}. Knowing whether a target device is running the expected software/firmware can be useful to the investigator, i.e., a malicious user may have modified the firmware. DL algorithms such as multi-layer perceptron (MLP) and long short-term memory (LSTM) have been used to detect anomalies in IoT devices through power consumption side-channels~\cite{wang2018deep}. Furthermore, various insights such as the identification of the specific hardware device or software application, and the behaviour of the software are shown to be identifiable with DL methods~\cite{laput2015sense,lerman2011side,callan2016zero,callan2016analyzing,nazari2017eddie,stone2016comparison}.

Ronald Rivest, one of the co-founders of the RSA algorithm, discussed the inter-relationship between cryptography and ML three decades ago~\cite{rivest1991cryptography}. Cryptanalysis attempts to retrieve cryptographic keys by analysing a large amount amount of information, i.e., plaintexts and ciphertexts, connected by an unknown key. There exists an interesting similarity between this and ML that eventually paved way to cryptographic key retrieval attacks leveraging ML and DL.

Template attacks are a common key retrieval approach whereby an attacker has a testing device similar to the target device~\cite{chari2002template}. A template can be built for the test device and subsequently used to attack the target device. It has been shown that SVMs are applicable in similar circumstances and provide comparable performance in key retrieval attacks~\cite{hospodar2011machine}. Experimental studies show that ML and DL methods can succeed even when cryptographic implementations use side-channel mitigation techniques to counter attacks~\cite{maghrebi2016breaking}. Furthermore, DL architectures, e.g., CNNs, are increasingly being applied to this problem~\cite{benadjila2018study}.

\subsubsection{Current Challenges and Future Directions}

It is reasonable to expect that a large percentage of computing devices encountered in digital forensic investigations in the future will be encrypted. Therefore, cryptography is turning into a critically important challenge in digital forensics. With the increasing popularity of software defined radio (SDR) hardware, acquisition of EM traces becomes easier and more affordable~\cite{machado2015software, bechet2019low}. Meanwhile, with the rapid increase of computational resources, ML and DL methods that are capable of performing key retrieval attacks can be expected to be more and more sophisticated. This can lead to a significant reduction in the time required for key retrieval.



Various side-channel mitigation techniques exist to defend against attacks to cryptographic implementations e.g., randomisation of operations, masking variables with random values, accessing critical variables indirectly via pointers, and hardware shielding~\cite{saputra2003masking, kim2016protecting, zankl2018side}. Furthermore, secure cores that are dedicated for cryptographic operations are increasingly present in modern computer processor chips. Operations performed inside such cores lower the side-channel information leakage, forcing attackers to use more sensitive measuring equipment and sophisticated pre-processing of EM traces in order to perform key retrieval attacks~\cite{fujino2017tamper, masure2020comprehensive}.


Software implementations of cryptographic algorithms tend to evolve over time due to updates carrying bug fixes and improvements. Such changes to software tend to impact the corresponding EM emission patterns. Therefore, ML and DL models that are trained to recognise patterns or retrieve keys can get affected by these changes. The ability of ML and DL models to generalise the minor changes of EM traces needs to be explored further. Meanwhile, tools and frameworks are needed to facilitate the application of ML/DL techniques for law-enforcement.

\subsection{Timeline/Event Reconstruction}

Event reconstruction in digital forensics has been defined in terms of finite state machines by~\citet{gladyshev2004finite}. However, it less formally refers to a process that can ``convert the state of the [digital] objects into the events that caused the state''~\cite{carrier2004event}. This can include simply being able to determine that some event occurred, or more precisely that an event occurred at a certain time. This second, more detailed event reconstruction would be achieved by looking at the timestamps recoverable from digital forensic artefacts. Sources of timestamps would include times from the file system, e.g., file modified, accessed, created, entry modified, etc., but can also include timestamps from inside more complex file formats, e.g., Windows Registry, SQLite databases, event logs, etc. 

The state of the art in terms of timestamp extraction is Plaso (log2timeline), which has many plugins and parsers\footnote{\url{https://plaso.readthedocs.io/en/latest/}}. However, the challenge is that the analysis of a system, even with minimal user activity, would generate millions of these timestamps. Attempts have been made to perform automated analysis of this high volume of timestamps and infer a usable activity history from this data. One approach by \citet{hargreaves2012automated} involved manually coding the pattern of low-level timestamps associated with a `higher level' event, e.g., a user opening a file on a Windows system produces a series of `low level' artefacts including entries in the Windows registry, link file changes, jump lists, and others. This can be manually encoded and pattern matched. However, this can be a time-consuming process to identify these changes, code and test them. It is also potentially error prone as subtle differences in behaviour of operating system versions could produce incorrect inferences. There are also representation problems for events, something that was examined by~\citet{chabot2014complete}, with a correlation of events also discussed in~\cite{chabot2015ontology}. 

\subsubsection{State of the Art of AI in Event Reconstruction}

Despite the potential of ML approaches in this area, there are relatively few papers on ML applied to pattern matching in timeline data. \citet{khan2007framework} and \citet{khan2012performance} discuss a neural network-based approach for event reconstruction using file system times and describe that neural networks are appropriate for dealing with the large volumes of data because of their parallelism and generalisation capabilities. They tested both feedforward and recurrent neural networks. \citet{turnbull2015automated} developed ParFor, which as a result of the explainability problems of other ML techniques, use Symbolic AI based on an ontological representation of forensic artefacts and implemented inferences such as computer on/off. However, \citet{studiawan2020sentiment} used DL techniques to highlight events of interest in a timeline based on positive or negative sentiment in the text-based representation of events (specifically operating system logs), e.g., `failed password' or `authentication failure'.

\subsubsection{Current Challenges and Future Directions}

There are a number of challenges in this area. Performing event reconstruction using timestamps inherently makes the assumption that the timestamps are correct. There are many reasons why this may not be the case, e.g., clock drift, manual changing of the system clock, overwritten timestamps as part of normal system processes, or anti-forensic techniques. There is some work in mitigating some of these, e.g., \citet{marrington2011cat} developed a rule-based approach to detecting timestamp inconsistencies, but there may be merit in testing a ML based approach to this problem too. It may also be possible to use many of the approaches developed in a network forensics and traffic analysis context and apply to artefact timeline analysis. Correct inference of user activity is also a challenge; the low-level events generated for one version of an operating system are not necessarily the same in other versions and evaluation of the correctness of the inferences is critical~\citet{jeyaraman2006empirical}. Finally, labelled and verified datasets for forensic timelines are very difficult to obtain and time consuming to generate at scale~\cite{du2020tracegen}. 

There are many areas to explore in the application of AI to event correlation. Aside from identification of individual events, it could also be possible to have higher level `anomaly detection' applied to a system. This is a very difficult problem given the multipurpose nature of typical computer systems and that the difference between legal and illegal activity may be very subtle. Nevertheless, in terms of applications, timelines have great potential to allow easier analysis across multiple applications, e.g., chat messages in multiple clients, or across devices~\cite{hargreaves2019synctriage}. Finally, a timeline-based view of activity is just one view of a data set, but it can provide a useful entry point into a dataset, allowing the view to be then `pivoted' to file system views, similar content and back to timelines.

\subsection{Multimedia Forensics}


Multimedia forensics is a branch of digital forensics that studies content such as audio, video and images that have been obtained as part of a digital forensics investigation and can include not just computers and mobile devices, but also CCTV analysis. There are a number of aspects to explore in this topic. The first is the problem of volume. Typical devices will contain thousands of media files and identifying those that are relevant can be a challenge as they cannot simply be keyword searched. The second area is analysis to determine the media's provenience, which could provide a link to a suspect. The third area is forgery detection as digital images can be easily tempered with. 

Object detection also has a role to play. A social media crowd sourcing approach by Europol has been used to trace objects to combat child abuse. The organisation explains that even the most innocent clues on photographs can aid investigations. Their aim is that once the origin of an object is identified, the LEA of the country involved will be informed to further investigate the lead and speed up the identification of both the offender and the victim.


\subsubsection{State of the Art of AI on Computer Vision}

In terms of identifying relevant images from a large set, the search for objects of interest in digital images is arduous due to the large volume of seized devices. The need for automated object detection, specially in low quality images is required and has also triggered the need to develop effective image mining systems for digital forensics purposes~\cite{brown2005}. 


In terms of general approaches, object identification can be tackled with CNNs. In 2016, \citet{grega2016automated} presented the automated detection and recognition of dangerous situations such as events where firearms and knives are present in CCTV footage. The algorithm proposed for knife detection is based on visual descriptors and ML. The algorithm for firearm detection is limited to a pistol and is based on a PCA approach. In 2017, a crowd-sourced and CV based approach to fight sex trafficking was proposed; hotel identification with a search-by-image based on features extracted from neural networks was implemented~\cite{stylianou2017traffickcam}. Later in 2019, \citet{xiao2019video} proposed a DL-based object detection and tracking algorithm to identify potential suspects from footage. Their approach for low quality video/image analysis is based on contrast limited adaptive histogram equalisation that improve CCTV quality and is used for Digital Forensic Investigations. Similarly, \citet{JASMINE2019833} proposed a real-time video quality enhancing method using adaptive histogram equalisation. Also, regarding the use of illicit substances, \citet{10.1145/3011871} proposed the tracking of drug dealing and abuse on the Instagram social network by using multi-modal analysis including methods such as multi-task learning and decision-level fusion; this approach enabled the ability to identify drug-related posts and the examination of behaviour patters of drug-related user accounts.

Specifically related to CSEM investigations, many practitioners are unfamiliar with AI, but demand automated nudity, age and skin tone detectors~\cite{sanchez2019practitioner}. This is unsurprising as it has been reported that some law enforcement personnel have suffered ill effects due to the continuous exposure of CSEM~\cite{wolak2009work}, and it has been proven to affect some groups by causing secondary traumatic stress disorder \cite{maceachern2011child, sanchez2019practitioner}. To lessen the exposure to CSEM, multiple approaches have been considered. Skin detection algorithms could potentially sift unnecessary images and flag inappropriate content. In 2005, \citet{ap2005algorithm} developed a skin colour distribution model based on RGB. The aforementioned nudity detection algorithm had a 95\% recall with a 5\% false positive rate. Later in 2016, Deep CNNs were used by \citet{NIAN2016283}. The latter demonstrates the advantage of using AI over hand-engineered visual features that are hard to analyse and select. The notable trend of CNNs has been flooding research topics in the past years.

Another relevant CV area is age estimation. In 2020,~\citet{anda2020UnderageAgeEstimation}, proposed the segregation of the age component from a CSEM investigative model. This approach tackles specifically the facial age estimation problem for underage subjects, which can be further consolidated with a nudity component to create a CSEM ensemble. This approach has achieved a mean absolute error (MAE) rate of 2.73 years. In order to tackle unbalanced dataset and bias, a balanced dataset generator was used~\cite{8424644}. Age estimation is a challenging task for both humans and computers. The range of factors that influence age prediction are considerable. Environment, habits, diets, use of anti-ageing products, smoking, drinking, drug abuse, skin tone, gender, etc. are only some identifiable parameters that can change the course of the appreciation of age. Nevertheless, in certain age groups (newborn and children), the influence of these factors has less of an impact. Age prediction may also have other applications for digital forensics including suspect and victim identification. Missing children cases could benefit from Generative Adversarial Networks that are able to estimate images of victims creating aged versions from an input image.


\subsubsection{State of the Art of AI in Forgery Detection}

Finally, as mentioned above, detecting forgeries in images is also a challenge. A digital image has been accepted as a ``proof of occurrence'' of an event \cite{kharrazi2004blind,sencar2009overview} and so it is important to demonstrate that it is authentic. \citet{farid2009image}, classified tools to detect image forgery into five categories:

\begin{enumerate}
    \item 
    Pixel-based techniques that detect statistical anomalies introduced at the pixel level.
    \item
    Format-based techniques that leverage the statistical correlations introduced by a specific lossy compression scheme.
    \item
    Camera-based techniques that exploit artefacts introduced by the camera lens, sensor, or on-chip post-processing.
    \item
    Physically based techniques that explicitly model and detect anomalies in the three-dimensional interaction between physical objects, light, and the camera.
    \item
    Geometric-based techniques that make measurements of objects in the world and their positions relative to the camera.
    
\end{enumerate}

The previous techniques are mainly statistically, geometrically and physically-based scientific methods, and are solely for the validation of the integrity of images. Nevertheless DL based techniques have also been used to detect image manipulation \cite{bayar2016deep,zhang2016image,rao2016deep,chen2015median}. The first three studies employ CNNs and the final study implements a Stacked Auto-encoder (SAE) approach.

\subsubsection{Current Challenges and Future Directions}

Automated mechanisms to detect CSEM have been used in the past with skin tone detection algorithms or hash comparisons. Nevertheless, either the performance has not been adequate or the approach used has been trivial. The rise of CNNs has enabled impressive and promising results. There are still a myriad of application to explore that could improve the performance of algorithms to detect CSEM. Images with low resolution and visibly challenging to the human eye could be tackled with the application of specific models trained on low quality data. Objects that have been found on CSEM with low quality can benefit with the creation of ensembles for different type of items matching certain quality standards.

However, the need for shared, well curated datasets in the research community is clear. Data pollution present in datasets that are already being shared in the community may present a risk to further research. Keeping big data under control may become a challenge and could be subject to data protection acts that would hinder certain types of longitudinal research. Unavailability of information due to ethical concerns and lack of transparency can impede the creation of reliable models. Non-robust models could also be subject to adversarial attacks that could bypass certain systems such as nudity detectors, and age limit systems.

Nevertheless, automation in multimedia forensics could help alleviate the digital forensic backlog by optimising analysis and prioritise artefacts in an intelligent manner. As previously highlighted, the usage of CNNs in digital forensics shows great promise. Proposed models should emphasise expertise while focusing on solving smaller problems rather than generalising to attempt to solve several problems. Ensemble models are considered to be more stable and, most importantly, predict better than single classifiers~\cite{lessmann2015benchmarking}; therefore, an ensemble of expert models would improve the performance while decreasing errors.

\subsection{Fingerprinting}
Device fingerprinting is a growing area of digital forensics. It ranges from server-side browser fingerprinting~\cite{eckersley2010unique} (based on the unique set of browser and extension metadata/configuration sent to a web server), camera sensor identification~\cite{lukas2006digital} (based on subtle imperfections of camera sensors), to malware behavioural analysis and classification~\cite{le2018deep} (based on program execution patterns). 

\subsubsection{State of the Art of AI in Fingerprinting}
The task of fingerprinting lends itself well to AI classification techniques. For example, malware classification has been a popular application area, with significant existing work in this area~\cite{gandotra2014malware}; both static~\cite{nath2014static,le2018deep} and dynamic analysis~\cite{kolosnjaji2016deep,huang2016mtnet}. With respect to the aforementioned media provenience issue, \citet{Tsai2007} were able to obtain highly accurate predictions with SVM on similar photographed scenes generated both by traditional and mobile-phone cameras. Also, CNNs have shown promising results on image recognition, video analysis and NLP. 

Similarly to how scratches on a bullet facilitate the identification of the weapon that shot it, subtle imperfections in digital camera sensors leave their imprint on the resultant digital photos and videos. This allows the subsequent association of this content with a specific camera sensor~\cite{lukas2006digital}. This approach can be used to identify both the specific make/model of the source camera, e.g., an iPhone 11, and potentially the specific camera, e.g., \textit{this} iPhone 11. Identifying the camera model with which a video has been taken can provide valuable insight in an investigation. \citet{FREIREOBREGON2018} implemented a Source Camera Identification (SCI) method that is able to infer the noise pattern of mobile camera sensors/fingerprints. Their CNN approach has achieved over a 90\% of accuracy in determining not only the brand of the phone but also identifying if the front or rear camera was used. CNNs are capable of performing image manipulation detection as well as camera model identification.

In a similar vein, fingerprinting techniques have also been used for authorship attribution. This authorship attribution ranges from open source intelligence/social media attribution~\cite{rocha2016authorship}, source code attribution~\cite{kalgutkar2019code} and malware attribution~\cite{alrabaee2016feasibility}. Authorship attribution relies on identifying unique programming or language traits of the individual behind the keyboard.

\subsubsection{Future Directions}

Device and user behavioural fingerprinting can greatly aid in anomaly detection. For networked devices, this can result in more accurate host-based and network-based intrusion detection. Modelling the usage/behavioural fingerprint of each user on a system can similarly be used as an indicator of account compromise. In online video streaming scenarios, camera sensor detection combined with device fingerprinting can be used to identify the source of the stream. 

\section{Conclusion}
This paper has shown how a range of AI techniques are currently used across different areas of digital forensics. It has also highlighted common challenges including availability of data sets in some areas, specific difficulties in explaining the results when certain techniques are used, and even challenges in releasing models where potentially restricted training data could be inferred from the models~\cite{oh2019towards}.
However, despite these challenges, there is enormous potential for future work. As discussed above, this is both in terms of improving the performance of some of the current techniques, but also that there are some approaches that have not yet been tested in individual areas. These gaps should now be more easily identified as a result of this systematisation of knowledge and help accelerate developments in this field.

\subsection{Future Directions}
The previous sections have shown that there is significant existing work in the application of AI to specific areas of digital forensics. This section discusses general challenges, and potential opportunities including unexplored areas where existing and emerging AI techniques have not yet been applied. 

In terms of general challenges, improving the accuracy of techniques is an obvious focus area. Specific to digital forensics, training models and measuring the accuracy is a challenge because of the lack of large, clean, labelled datasets in some areas or existing datasets not being publicly available. While there are extensive datasets that can be used to train computer vision based approaches, the availability of sensitive datasets, such as CSEM datasets, are understandably and necessarily restricted. Datasets for whole hard disk approaches, e.g., needed for timeline analysis, do not exist in a useful manner; where user activity is clearly documented and labelled allowing DL techniques to identify relevant features. Even if such disk images were produced, having sufficient background `noise' is also difficult, meaning that techniques developed in a research setting are unlikely to work when exposed to data from a real investigation. Automated digital `story' generation is needed to address these issues.

While Explainable AI is a general computer science problem, for digital forensics this is paramount for the court admissibility and understanding of evidence. However, it should be highlighted that there is some subtlety to this. For example, an AI process reporting that a system containing criminal activity needs to be able to produce a very clear explanation of why that is the case. However, a human-in-the-loop approach that is designed to highlight to an investigator data that is likely to be relevant does not necessarily have the same explainability requirement and is something called for by law enforcement~\cite{sanchez2019practitioner}. There is still a danger here in that bias can be a problem in investigations in general~\cite{meissner2002he}, but a system that is promoting `relevant evidence' has the potential to bias an investigator. A related problem is validation -- increasingly necessary for methods used in a digital forensic context. This means that a technique should be applied to known data and produce an expected result. Subsequently, once validated, that technique can be used. New versions of software, or in the case of AI models, new models mean the techniques should be re-validated. The edge case is in a scenario where an AI model is updating live, for example learning from on-going case processing to expedite evidence discovery in future cases. In this case, the result the technique may produce may change on a daily basis, posing a significant validation challenge. 

Finally, it should also be considered whether sharing models is appropriate in some contexts. AI trained models in the context of GDPR and summary of attacks such as `model inversion' and `membership inference' is discussed by~\citet{veale2018algorithms}. This is therefore worth considering when developing digital forensic solutions using AI and potentially sensitive training data. 

Despite these challenges, there are many opportunities to enhance AI applications and to apply AI to additional areas of digital forensics. These include inference of behaviour from data obtained from novel sources including smart homes, IoT sensors, vehicle forensics, and combinations thereof. Indeed AI techniques could potentially assist any time there is a need to correlate data from multiple sources, either from multiple suspects, devices or cases. Non-AI based efforts such as standard form of representations, e.g., CASE~\cite{casey2018evolution} will be critical for such efforts. 

There will also be significant opportunities in the future for the investigation of AI based systems themselves. Determining the cause of a decision made by a self-driving car, a smart building, or a SCADA system, will be a new area for digital forensics, although the concept is discussed by~\citet{schneiderai}. The investigation of these systems will require significant effort on behalf of the investigator in terms of understanding the models, their training data, and the state of the model's inputs when the decision was made. This will also require a reasonable level of explainable AI. Of course, the investigation of digital forensic AI systems themselves will be far from exempt from this scrutiny.

\bibliographystyle{ACM-Reference-Format}
\bibliography{sample-base}


\begin{thebibliography}{124}


\ifx \showCODEN    \undefined \def \showCODEN     #1{\unskip}     \fi
\ifx \showDOI      \undefined \def \showDOI       #1{#1}\fi
\ifx \showISBNx    \undefined \def \showISBNx     #1{\unskip}     \fi
\ifx \showISBNxiii \undefined \def \showISBNxiii  #1{\unskip}     \fi
\ifx \showISSN     \undefined \def \showISSN      #1{\unskip}     \fi
\ifx \showLCCN     \undefined \def \showLCCN      #1{\unskip}     \fi
\ifx \shownote     \undefined \def \shownote      #1{#1}          \fi
\ifx \showarticletitle \undefined \def \showarticletitle #1{#1}   \fi
\ifx \showURL      \undefined \def \showURL       {\relax}        \fi
\providecommand\bibfield[2]{#2}
\providecommand\bibinfo[2]{#2}
\providecommand\natexlab[1]{#1}
\providecommand\showeprint[2][]{arXiv:#2}

\bibitem[\protect\citeauthoryear{Abdulhammed, Musafer, Alessa, Faezipour, and
  Abuzneid}{Abdulhammed et~al\mbox{.}}{2019}]%
        {Abdulhammed:2019}
\bibfield{author}{\bibinfo{person}{Razan Abdulhammed}, \bibinfo{person}{Hassan
  Musafer}, \bibinfo{person}{Ali Alessa}, \bibinfo{person}{Miad Faezipour},
  {and} \bibinfo{person}{Abdelshakour Abuzneid}.}
  \bibinfo{year}{2019}\natexlab{}.
\newblock \showarticletitle{Features Dimensionality Reduction Approaches for
  Machine Learning Based Network Intrusion Detection}.
\newblock  \bibinfo{number}{8} (\bibinfo{date}{March} \bibinfo{year}{2019}).
\newblock


\bibitem[\protect\citeauthoryear{Ahmed, Naser, and Hu}{Ahmed
  et~al\mbox{.}}{2016}]%
        {Ahmed:2016}
\bibfield{author}{\bibinfo{person}{Mohiuddin Ahmed}, \bibinfo{person}{Abdun.
  Naser}, {and} \bibinfo{person}{Jiankun Hu}.} \bibinfo{year}{2016}\natexlab{}.
\newblock \showarticletitle{A Survey of Network Anomaly Detection Techniques}.
\newblock \bibinfo{journal}{\emph{J. Netw. Comput. Appl.}}
  \bibinfo{volume}{60}, \bibinfo{number}{C} (\bibinfo{date}{Jan.}
  \bibinfo{year}{2016}), \bibinfo{pages}{19--31}.
\newblock
\showISSN{1084-8045}
\urldef\tempurl%
\url{https://doi.org/10.1016/j.jnca.2015.11.016}
\showDOI{\tempurl}


\bibitem[\protect\citeauthoryear{Aksu, Ustebay, Aydin, and Atmaca}{Aksu
  et~al\mbox{.}}{2018}]%
        {Aksu:2018b}
\bibfield{author}{\bibinfo{person}{Dogukan Aksu}, \bibinfo{person}{Serpil
  Ustebay}, \bibinfo{person}{Muhammed~Ali Aydin}, {and}
  \bibinfo{person}{T{\"u}lin Atmaca}.} \bibinfo{year}{2018}\natexlab{}.
\newblock \bibinfo{booktitle}{\emph{Intrusion Detection with Comparative
  Analysis of Supervised Learning Techniques and Fisher Score Feature Selection
  Algorithm}}.
\newblock \bibinfo{pages}{141--149}.
\newblock
\urldef\tempurl%
\url{https://doi.org/10.1007/978-3-030-00840-6_16}
\showDOI{\tempurl}


\bibitem[\protect\citeauthoryear{Alauthman, Aslam, Al-kasassbeh, Khan,
  Al-Qerem, and Choo]}{Alauthman et~al\mbox{.}}{2020}]%
        {ALAUTHMAN:2000}
\bibfield{author}{\bibinfo{person}{Mohammad Alauthman}, \bibinfo{person}{Nauman
  Aslam}, \bibinfo{person}{Mouhammd Al-kasassbeh}, \bibinfo{person}{Suleman
  Khan}, \bibinfo{person}{Ahmad Al-Qerem}, {and}
  \bibinfo{person}{Kim-Kwang~[Raymond Choo]}.} \bibinfo{year}{2020}\natexlab{}.
\newblock \showarticletitle{An efficient reinforcement learning-based Botnet
  detection approach}.
\newblock \bibinfo{journal}{\emph{Journal of Network and Computer
  Applications}}  \bibinfo{volume}{150} (\bibinfo{year}{2020}),
  \bibinfo{pages}{102479}.
\newblock
\showISSN{1084-8045}
\urldef\tempurl%
\url{https://doi.org/10.1016/j.jnca.2019.102479}
\showDOI{\tempurl}


\bibitem[\protect\citeauthoryear{Alrabaee, Shirani, Debbabi, and Wang}{Alrabaee
  et~al\mbox{.}}{2016}]%
        {alrabaee2016feasibility}
\bibfield{author}{\bibinfo{person}{Saed Alrabaee}, \bibinfo{person}{Paria
  Shirani}, \bibinfo{person}{Mourad Debbabi}, {and} \bibinfo{person}{Lingyu
  Wang}.} \bibinfo{year}{2016}\natexlab{}.
\newblock \showarticletitle{On the feasibility of malware authorship
  attribution}. In \bibinfo{booktitle}{\emph{International Symposium on
  Foundations and Practice of Security}}. Springer, \bibinfo{pages}{256--272}.
\newblock


\bibitem[\protect\citeauthoryear{Amouri, Alaparthy, and Morgera}{Amouri
  et~al\mbox{.}}{2018}]%
        {Amouri:2018}
\bibfield{author}{\bibinfo{person}{Amar Amouri}, \bibinfo{person}{Vishwa
  Alaparthy}, {and} \bibinfo{person}{Salvatore~Dominic Morgera}.}
  \bibinfo{year}{2018}\natexlab{}.
\newblock \showarticletitle{Cross layer-based intrusion detection based on
  network behavior for IoT}. In \bibinfo{booktitle}{\emph{2018 IEEE 19th
  Wireless and Microwave Technology Conference (WAMICON)}}.
  \bibinfo{pages}{1--4}.
\newblock
\urldef\tempurl%
\url{https://doi.org/10.1109/WAMICON.2018.8363921}
\showDOI{\tempurl}


\bibitem[\protect\citeauthoryear{Anda, Le-Khac, and Scanlon}{Anda
  et~al\mbox{.}}{2020}]%
        {anda2020UnderageAgeEstimation}
\bibfield{author}{\bibinfo{person}{Felix Anda}, \bibinfo{person}{Nhien-An
  Le-Khac}, {and} \bibinfo{person}{Mark Scanlon}.}
  \bibinfo{year}{2020}\natexlab{}.
\newblock \showarticletitle{{DeepUAge: Improving Underage Age Estimation
  Accuracy to Aid CSEM Investigation}}.
\newblock \bibinfo{journal}{\emph{{Forensic Science International: Digital
  Investigation}}}  \bibinfo{volume}{32} (\bibinfo{date}{04}
  \bibinfo{year}{2020}), \bibinfo{pages}{300921}.
\newblock
\showISSN{2666-2817}
\urldef\tempurl%
\url{https://doi.org/10.1016/j.fsidi.2020.300921}
\showDOI{\tempurl}


\bibitem[\protect\citeauthoryear{Ap-Apid}{Ap-Apid}{2005}]%
        {ap2005algorithm}
\bibfield{author}{\bibinfo{person}{Rigan Ap-Apid}.}
  \bibinfo{year}{2005}\natexlab{}.
\newblock \showarticletitle{An algorithm for nudity detection}. In
  \bibinfo{booktitle}{\emph{5th Philippine Computing Science Congress}}.
  \bibinfo{pages}{201--205}.
\newblock


\bibitem[\protect\citeauthoryear{Bayar and Stamm}{Bayar and Stamm}{2016}]%
        {bayar2016deep}
\bibfield{author}{\bibinfo{person}{Belhassen Bayar} {and}
  \bibinfo{person}{Matthew~C Stamm}.} \bibinfo{year}{2016}\natexlab{}.
\newblock \showarticletitle{A deep learning approach to universal image
  manipulation detection using a new convolutional layer}. In
  \bibinfo{booktitle}{\emph{Proceedings of the 4th ACM Workshop on Information
  Hiding and Multimedia Security}}. ACM, \bibinfo{pages}{5--10}.
\newblock


\bibitem[\protect\citeauthoryear{Bechet, Helbet, Bouleanu, Sarbu, Miclaus, and
  Bechet}{Bechet et~al\mbox{.}}{2019}]%
        {bechet2019low}
\bibfield{author}{\bibinfo{person}{Andrei~Cristian Bechet},
  \bibinfo{person}{Robert Helbet}, \bibinfo{person}{Iulian Bouleanu},
  \bibinfo{person}{Annamaria Sarbu}, \bibinfo{person}{Simona Miclaus}, {and}
  \bibinfo{person}{Paul Bechet}.} \bibinfo{year}{2019}\natexlab{}.
\newblock \showarticletitle{Low Cost Solution Based on Software Defined Radio
  for the RF Exposure Assessment: A Performance Analysis}. In
  \bibinfo{booktitle}{\emph{2019 11th International Symposium on Advanced
  Topics in Electrical Engineering (ATEE)}}. IEEE, \bibinfo{pages}{1--4}.
\newblock


\bibitem[\protect\citeauthoryear{Benadjila, Prouff, Strullu, Cagli, and
  Dumas}{Benadjila et~al\mbox{.}}{2018}]%
        {benadjila2018study}
\bibfield{author}{\bibinfo{person}{Ryad Benadjila}, \bibinfo{person}{Emmanuel
  Prouff}, \bibinfo{person}{R{\'e}mi Strullu}, \bibinfo{person}{Eleonora
  Cagli}, {and} \bibinfo{person}{C{\'e}cile Dumas}.}
  \bibinfo{year}{2018}\natexlab{}.
\newblock \showarticletitle{Study of deep learning techniques for side-channel
  analysis and introduction to ASCAD database}.
\newblock \bibinfo{journal}{\emph{ANSSI, France \& CEA, LETI, MINATEC Campus,
  France. Online verf{\"u}gbar unter https://eprint. iacr. org/2018/053. pdf,
  zuletzt gepr{\"u}ft am}}  \bibinfo{volume}{22} (\bibinfo{year}{2018}),
  \bibinfo{pages}{2018}.
\newblock


\bibitem[\protect\citeauthoryear{Biradar and Padmavathi}{Biradar and
  Padmavathi}{2020}]%
        {Biradar:2020}
\bibfield{author}{\bibinfo{person}{Anuradha~D. Biradar} {and}
  \bibinfo{person}{B. Padmavathi}.} \bibinfo{year}{2020}\natexlab{}.
\newblock \showarticletitle{BotHook: A Supervised Machine Learning Approach for
  Botnet Detection Using DNS Query Data}. In \bibinfo{booktitle}{\emph{ICCCE
  2019}}, \bibfield{editor}{\bibinfo{person}{Amit Kumar} {and}
  \bibinfo{person}{Stefan Mozar}} (Eds.). \bibinfo{publisher}{Springer
  Singapore}, \bibinfo{address}{Singapore}, \bibinfo{pages}{261--269}.
\newblock
\showISBNx{978-981-13-8715-9}


\bibitem[\protect\citeauthoryear{Brown, Pham, and Vel}{Brown
  et~al\mbox{.}}{2005}]%
        {brown2005}
\bibfield{author}{\bibinfo{person}{Ross Brown}, \bibinfo{person}{Binh Pham},
  {and} \bibinfo{person}{Olivier Vel}.} \bibinfo{year}{2005}\natexlab{}.
\newblock \showarticletitle{Design of a Digital Forensics Image Mining System}.
\newblock \bibinfo{journal}{\emph{Lecture Notes in Computer Science}}.
\newblock
\urldef\tempurl%
\url{https://doi.org/10.1007/11553939_57}
\showDOI{\tempurl}


\bibitem[\protect\citeauthoryear{Buczak and Guven}{Buczak and Guven}{2016}]%
        {Buczak:2016}
\bibfield{author}{\bibinfo{person}{Anna~L. Buczak} {and} \bibinfo{person}{Erhan
  Guven}.} \bibinfo{year}{2016}\natexlab{}.
\newblock \showarticletitle{A Survey of Data Mining and Machine Learning
  Methods for Cyber Security Intrusion Detection}.
\newblock \bibinfo{journal}{\emph{{IEEE} Communications Surveys and Tutorials}}
  \bibinfo{volume}{18}, \bibinfo{number}{2} (\bibinfo{year}{2016}),
  \bibinfo{pages}{1153--1176}.
\newblock
\urldef\tempurl%
\url{https://doi.org/10.1109/COMST.2015.2494502}
\showDOI{\tempurl}


\bibitem[\protect\citeauthoryear{Callan, Behrang, Zajic, Prvulovic, and
  Orso}{Callan et~al\mbox{.}}{2016}]%
        {callan2016zero}
\bibfield{author}{\bibinfo{person}{Robert Callan}, \bibinfo{person}{Farnaz
  Behrang}, \bibinfo{person}{Alenka Zajic}, \bibinfo{person}{Milos Prvulovic},
  {and} \bibinfo{person}{Alessandro Orso}.} \bibinfo{year}{2016}\natexlab{}.
\newblock \showarticletitle{Zero-overhead profiling via em emanations}. In
  \bibinfo{booktitle}{\emph{Proceedings of the 25th International Symposium on
  Software Testing and Analysis}}. ACM, \bibinfo{pages}{401--412}.
\newblock


\bibitem[\protect\citeauthoryear{Callan}{Callan}{2016}]%
        {callan2016analyzing}
\bibfield{author}{\bibinfo{person}{Robert~Locke Callan}.}
  \bibinfo{year}{2016}\natexlab{}.
\newblock \emph{\bibinfo{title}{Analyzing software using unintentional
  electromagnetic emanations from computing devices}}.
\newblock \bibinfo{thesistype}{Ph.D. Dissertation}. \bibinfo{school}{Georgia
  Institute of Technology}.
\newblock


\bibitem[\protect\citeauthoryear{Carrier and Spafford}{Carrier and
  Spafford}{2004}]%
        {carrier2004event}
\bibfield{author}{\bibinfo{person}{Brian Carrier} {and}
  \bibinfo{person}{Eugene~H Spafford}.} \bibinfo{year}{2004}\natexlab{}.
\newblock \showarticletitle{An event-based digital forensic investigation
  framework}. In \bibinfo{booktitle}{\emph{Digital forensic research
  workshop}}. \bibinfo{pages}{11--13}.
\newblock


\bibitem[\protect\citeauthoryear{Casey, Barnum, Griffith, Snyder, van Beek, and
  Nelson}{Casey et~al\mbox{.}}{2018}]%
        {casey2018evolution}
\bibfield{author}{\bibinfo{person}{Eoghan Casey}, \bibinfo{person}{Sean
  Barnum}, \bibinfo{person}{Ryan Griffith}, \bibinfo{person}{Jonathan Snyder},
  \bibinfo{person}{Harm van Beek}, {and} \bibinfo{person}{Alex Nelson}.}
  \bibinfo{year}{2018}\natexlab{}.
\newblock \showarticletitle{The evolution of expressing and exchanging
  cyber-investigation information in a standardized form}.
\newblock In \bibinfo{booktitle}{\emph{Handling and Exchanging Electronic
  Evidence Across Europe}}. \bibinfo{publisher}{Springer},
  \bibinfo{pages}{43--58}.
\newblock


\bibitem[\protect\citeauthoryear{Chabot, Bertaux, Nicolle, and Kechadi}{Chabot
  et~al\mbox{.}}{2014}]%
        {chabot2014complete}
\bibfield{author}{\bibinfo{person}{Yoan Chabot}, \bibinfo{person}{Aur{\'e}lie
  Bertaux}, \bibinfo{person}{Christophe Nicolle}, {and}
  \bibinfo{person}{M-Tahar Kechadi}.} \bibinfo{year}{2014}\natexlab{}.
\newblock \showarticletitle{A complete formalized knowledge representation
  model for advanced digital forensics timeline analysis}.
\newblock \bibinfo{journal}{\emph{Digital Investigation}}  \bibinfo{volume}{11}
  (\bibinfo{year}{2014}), \bibinfo{pages}{S95--S105}.
\newblock


\bibitem[\protect\citeauthoryear{Chabot, Bertaux, Nicolle, and Kechadi}{Chabot
  et~al\mbox{.}}{2015}]%
        {chabot2015ontology}
\bibfield{author}{\bibinfo{person}{Yoan Chabot}, \bibinfo{person}{Aur{\'e}lie
  Bertaux}, \bibinfo{person}{Christophe Nicolle}, {and} \bibinfo{person}{Tahar
  Kechadi}.} \bibinfo{year}{2015}\natexlab{}.
\newblock \showarticletitle{An ontology-based approach for the reconstruction
  and analysis of digital incidents timelines}.
\newblock \bibinfo{journal}{\emph{Digital Investigation}}  \bibinfo{volume}{15}
  (\bibinfo{year}{2015}), \bibinfo{pages}{83--100}.
\newblock


\bibitem[\protect\citeauthoryear{Chari, Rao, and Rohatgi}{Chari
  et~al\mbox{.}}{2002}]%
        {chari2002template}
\bibfield{author}{\bibinfo{person}{Suresh Chari}, \bibinfo{person}{Josyula~R
  Rao}, {and} \bibinfo{person}{Pankaj Rohatgi}.}
  \bibinfo{year}{2002}\natexlab{}.
\newblock \showarticletitle{Template attacks}. In
  \bibinfo{booktitle}{\emph{International Workshop on Cryptographic Hardware
  and Embedded Systems}}. Springer, \bibinfo{pages}{13--28}.
\newblock


\bibitem[\protect\citeauthoryear{Chen, Kang, Liu, and Wang}{Chen
  et~al\mbox{.}}{2015}]%
        {chen2015median}
\bibfield{author}{\bibinfo{person}{Jiansheng Chen}, \bibinfo{person}{Xiangui
  Kang}, \bibinfo{person}{Ye Liu}, {and} \bibinfo{person}{Z~Jane Wang}.}
  \bibinfo{year}{2015}\natexlab{}.
\newblock \showarticletitle{Median filtering forensics based on convolutional
  neural networks}.
\newblock \bibinfo{journal}{\emph{IEEE Signal Processing Letters}}
  \bibinfo{volume}{22}, \bibinfo{number}{11} (\bibinfo{year}{2015}),
  \bibinfo{pages}{1849--1853}.
\newblock


\bibitem[\protect\citeauthoryear{Chen, Liao, Jiang, Fang, Yiu, Xi, Li, Yi,
  Wang, Hui, et~al\mbox{.}}{Chen et~al\mbox{.}}{2018}]%
        {chen2018file}
\bibfield{author}{\bibinfo{person}{Qian Chen}, \bibinfo{person}{Qing Liao},
  \bibinfo{person}{Zoe~L Jiang}, \bibinfo{person}{Junbin Fang},
  \bibinfo{person}{Siuming Yiu}, \bibinfo{person}{Guikai Xi},
  \bibinfo{person}{Rong Li}, \bibinfo{person}{Zhengzhong Yi},
  \bibinfo{person}{Xuan Wang}, \bibinfo{person}{Lucas~CK Hui}, {et~al\mbox{.}}}
  \bibinfo{year}{2018}\natexlab{}.
\newblock \showarticletitle{File fragment classification using grayscale image
  conversion and deep learning in digital forensics}. In
  \bibinfo{booktitle}{\emph{2018 IEEE Security and Privacy Workshops (SPW)}}.
  IEEE, \bibinfo{pages}{140--147}.
\newblock


\bibitem[\protect\citeauthoryear{Deng, Li, Yao, Cox, and Wang}{Deng
  et~al\mbox{.}}{2018}]%
        {Deng:2018}
\bibfield{author}{\bibinfo{person}{Lianbing Deng}, \bibinfo{person}{Daming Li},
  \bibinfo{person}{Xiang Yao}, \bibinfo{person}{David Cox}, {and}
  \bibinfo{person}{Haixiang Wang}.} \bibinfo{year}{2018}\natexlab{}.
\newblock \showarticletitle{Mobile network intrusion detection for IoT system
  based on transfer learning algorithm}.
\newblock \bibinfo{journal}{\emph{Cluster Computing}} (\bibinfo{date}{31 Jan}
  \bibinfo{year}{2018}).
\newblock
\showISSN{1573-7543}
\urldef\tempurl%
\url{https://doi.org/10.1007/s10586-018-1847-2}
\showDOI{\tempurl}


\bibitem[\protect\citeauthoryear{Du, Hargreaves, Sheppard, and Scanlon}{Du
  et~al\mbox{.}}{2020}]%
        {du2020tracegen}
\bibfield{author}{\bibinfo{person}{Xiaoyu Du}, \bibinfo{person}{Christopher
  Hargreaves}, \bibinfo{person}{John Sheppard}, {and} \bibinfo{person}{Mark
  Scanlon}.} \bibinfo{year}{2020}\natexlab{}.
\newblock \showarticletitle{{TraceGen: User Activity Emulation for Digital
  Forensic Test Image Generation}}.
\newblock \bibinfo{journal}{\emph{{Forensic Science International: Digital
  Investigation}}} (\bibinfo{date}{09} \bibinfo{year}{2020}).
\newblock
\showISSN{2666-2817}
\newblock
\shownote{Proceedings of DFRWS APAC 2020.}


\bibitem[\protect\citeauthoryear{Du, Le-Khac, and Scanlon}{Du
  et~al\mbox{.}}{2017}]%
        {du2017processmodelsdfaas}
\bibfield{author}{\bibinfo{person}{Xiaoyu Du}, \bibinfo{person}{Nhien-An
  Le-Khac}, {and} \bibinfo{person}{Mark Scanlon}.}
  \bibinfo{year}{2017}\natexlab{}.
\newblock \showarticletitle{{Evaluation of Digital Forensic Process Models with
  Respect to Digital Forensics as a Service}}. In
  \bibinfo{booktitle}{\emph{{Proceedings of the 16th European Conference on
  Cyber Warfare and Security (ECCWS 2017)}}}. \bibinfo{publisher}{ACPI},
  \bibinfo{address}{Dublin, Ireland}, \bibinfo{pages}{573--581}.
\newblock


\bibitem[\protect\citeauthoryear{Eckersley}{Eckersley}{2010}]%
        {eckersley2010unique}
\bibfield{author}{\bibinfo{person}{Peter Eckersley}.}
  \bibinfo{year}{2010}\natexlab{}.
\newblock \showarticletitle{How unique is your web browser?}. In
  \bibinfo{booktitle}{\emph{International Symposium on Privacy Enhancing
  Technologies Symposium}}. Springer, \bibinfo{pages}{1--18}.
\newblock


\bibitem[\protect\citeauthoryear{Elrawy, Awad, and Hamed}{Elrawy
  et~al\mbox{.}}{2018}]%
        {Elrawy:2018}
\bibfield{author}{\bibinfo{person}{Mohamed~Faisal Elrawy},
  \bibinfo{person}{Ali~Ismail Awad}, {and} \bibinfo{person}{Hesham F.~A.
  Hamed}.} \bibinfo{year}{2018}\natexlab{}.
\newblock \showarticletitle{Intrusion detection systems for IoT-based smart
  environments: a survey}.
\newblock \bibinfo{journal}{\emph{Journal of Cloud Computing}}
  \bibinfo{volume}{7}, \bibinfo{number}{1} (\bibinfo{date}{04 Dec}
  \bibinfo{year}{2018}), \bibinfo{pages}{21}.
\newblock
\showISSN{2192-113X}
\urldef\tempurl%
\url{https://doi.org/10.1186/s13677-018-0123-6}
\showDOI{\tempurl}


\bibitem[\protect\citeauthoryear{EURPOL}{EURPOL}{2019}]%
        {Europol01}
\bibfield{author}{\bibinfo{person}{EURPOL}.} \bibinfo{year}{2019}\natexlab{}.
\newblock \bibinfo{booktitle}{\emph{Global Guidelines for Digital Forensic
  Laboratories}}.
\newblock
\urldef\tempurl%
\url{https://www.interpol.int/content/download/13501/file/INTERPOL_DFL_GlobalGuidelinesDigitalForensicsLaboratory.pdf}
\showURL{%
\tempurl}


\bibitem[\protect\citeauthoryear{Farid}{Farid}{2009}]%
        {farid2009image}
\bibfield{author}{\bibinfo{person}{Hany Farid}.}
  \bibinfo{year}{2009}\natexlab{}.
\newblock \showarticletitle{Image forgery detection}.
\newblock \bibinfo{journal}{\emph{IEEE Signal processing magazine}}
  \bibinfo{volume}{26}, \bibinfo{number}{2} (\bibinfo{year}{2009}),
  \bibinfo{pages}{16--25}.
\newblock


\bibitem[\protect\citeauthoryear{Felix~Anda and {Scanlon}}{Felix~Anda and
  {Scanlon}}{2018}]%
        {8424644}
\bibfield{author}{\bibinfo{person}{Nhien-An Le-Khac Felix~Anda, David~Lillis}
  {and} \bibinfo{person}{Mark {Scanlon}}.} \bibinfo{year}{2018}\natexlab{}.
\newblock \showarticletitle{Evaluating Automated Facial Age Estimation
  Techniques for Digital Forensics}. In \bibinfo{booktitle}{\emph{2018 IEEE
  Security and Privacy Workshops (SPW)}}. \bibinfo{pages}{129--139}.
\newblock


\bibitem[\protect\citeauthoryear{Fitzgerald, Mathews, Morris, and
  Zhulyn}{Fitzgerald et~al\mbox{.}}{2012}]%
        {fitzgerald2012using}
\bibfield{author}{\bibinfo{person}{Simran Fitzgerald}, \bibinfo{person}{George
  Mathews}, \bibinfo{person}{Colin Morris}, {and} \bibinfo{person}{Oles
  Zhulyn}.} \bibinfo{year}{2012}\natexlab{}.
\newblock \showarticletitle{Using NLP techniques for file fragment
  classification}.
\newblock \bibinfo{journal}{\emph{Digital Investigation}}  \bibinfo{volume}{9}
  (\bibinfo{year}{2012}), \bibinfo{pages}{S44--S49}.
\newblock


\bibitem[\protect\citeauthoryear{Flach}{Flach}{2012}]%
        {flach2012machine}
\bibfield{author}{\bibinfo{person}{Peter Flach}.}
  \bibinfo{year}{2012}\natexlab{}.
\newblock \bibinfo{booktitle}{\emph{Machine learning: the art and science of
  algorithms that make sense of data}}.
\newblock \bibinfo{publisher}{Cambridge University Press}.
\newblock


\bibitem[\protect\citeauthoryear{Freire-Obregon, Narducci, Barra, and
  Castrillon-Santana}{Freire-Obregon et~al\mbox{.}}{2018}]%
        {FREIREOBREGON2018}
\bibfield{author}{\bibinfo{person}{David Freire-Obregon},
  \bibinfo{person}{Fabio Narducci}, \bibinfo{person}{Silvio Barra}, {and}
  \bibinfo{person}{Modesto Castrillon-Santana}.}
  \bibinfo{year}{2018}\natexlab{}.
\newblock \showarticletitle{Deep learning for source camera identification on
  mobile devices}.
\newblock \bibinfo{journal}{\emph{Pattern Recognition Letters}}
  (\bibinfo{year}{2018}).
\newblock
\showISSN{0167-8655}
\urldef\tempurl%
\url{https://doi.org/10.1016/j.patrec.2018.01.005}
\showDOI{\tempurl}


\bibitem[\protect\citeauthoryear{Fujino, Kubota, and Shiozaki}{Fujino
  et~al\mbox{.}}{2017}]%
        {fujino2017tamper}
\bibfield{author}{\bibinfo{person}{Takeshi Fujino}, \bibinfo{person}{Takaya
  Kubota}, {and} \bibinfo{person}{Mitsuru Shiozaki}.}
  \bibinfo{year}{2017}\natexlab{}.
\newblock \showarticletitle{Tamper-resistant cryptographic hardware}.
\newblock \bibinfo{journal}{\emph{IEICE Electronics Express}}
  \bibinfo{volume}{14}, \bibinfo{number}{2} (\bibinfo{year}{2017}),
  \bibinfo{pages}{20162004--20162004}.
\newblock


\bibitem[\protect\citeauthoryear{Gandotra, Bansal, and Sofat}{Gandotra
  et~al\mbox{.}}{2014}]%
        {gandotra2014malware}
\bibfield{author}{\bibinfo{person}{Ekta Gandotra}, \bibinfo{person}{Divya
  Bansal}, {and} \bibinfo{person}{Sanjeev Sofat}.}
  \bibinfo{year}{2014}\natexlab{}.
\newblock \showarticletitle{Malware analysis and classification: A survey}.
\newblock \bibinfo{journal}{\emph{Journal of Information Security}}
  \bibinfo{volume}{2014} (\bibinfo{year}{2014}).
\newblock


\bibitem[\protect\citeauthoryear{Garfinkel, Farrell, Roussev, and
  Dinolt}{Garfinkel et~al\mbox{.}}{2009}]%
        {garfinkel2009bringing}
\bibfield{author}{\bibinfo{person}{Simson Garfinkel}, \bibinfo{person}{Paul
  Farrell}, \bibinfo{person}{Vassil Roussev}, {and} \bibinfo{person}{George
  Dinolt}.} \bibinfo{year}{2009}\natexlab{}.
\newblock \showarticletitle{Bringing science to digital forensics with
  standardized forensic corpora}.
\newblock \bibinfo{journal}{\emph{digital investigation}}  \bibinfo{volume}{6}
  (\bibinfo{year}{2009}), \bibinfo{pages}{S2--S11}.
\newblock


\bibitem[\protect\citeauthoryear{Garfinkel}{Garfinkel}{2007}]%
        {garfinkel2007carving}
\bibfield{author}{\bibinfo{person}{Simson~L Garfinkel}.}
  \bibinfo{year}{2007}\natexlab{}.
\newblock \showarticletitle{Carving contiguous and fragmented files with fast
  object validation}.
\newblock \bibinfo{journal}{\emph{digital investigation}}  \bibinfo{volume}{4}
  (\bibinfo{year}{2007}), \bibinfo{pages}{2--12}.
\newblock


\bibitem[\protect\citeauthoryear{Garfinkel, Parker-Wood, Huynh, and
  Migletz}{Garfinkel et~al\mbox{.}}{2010}]%
        {garfinkel2010automated}
\bibfield{author}{\bibinfo{person}{Simson~L Garfinkel},
  \bibinfo{person}{Aleatha Parker-Wood}, \bibinfo{person}{Daniel Huynh}, {and}
  \bibinfo{person}{James Migletz}.} \bibinfo{year}{2010}\natexlab{}.
\newblock \showarticletitle{An automated solution to the multiuser carved data
  ascription problem}.
\newblock \bibinfo{journal}{\emph{IEEE Transactions on Information Forensics
  and Security}} \bibinfo{volume}{5}, \bibinfo{number}{4}
  (\bibinfo{year}{2010}), \bibinfo{pages}{868--882}.
\newblock


\bibitem[\protect\citeauthoryear{Gladyshev and Patel}{Gladyshev and
  Patel}{2004}]%
        {gladyshev2004finite}
\bibfield{author}{\bibinfo{person}{Pavel Gladyshev} {and}
  \bibinfo{person}{Ahmed Patel}.} \bibinfo{year}{2004}\natexlab{}.
\newblock \showarticletitle{Finite state machine approach to digital event
  reconstruction}.
\newblock \bibinfo{journal}{\emph{Digital Investigation}} \bibinfo{volume}{1},
  \bibinfo{number}{2} (\bibinfo{year}{2004}), \bibinfo{pages}{130--149}.
\newblock


\bibitem[\protect\citeauthoryear{Grajeda, Breitinger, and Baggili}{Grajeda
  et~al\mbox{.}}{2017}]%
        {grajeda2017availability}
\bibfield{author}{\bibinfo{person}{Cinthya Grajeda}, \bibinfo{person}{Frank
  Breitinger}, {and} \bibinfo{person}{Ibrahim Baggili}.}
  \bibinfo{year}{2017}\natexlab{}.
\newblock \showarticletitle{Availability of datasets for digital forensics--and
  what is missing}.
\newblock \bibinfo{journal}{\emph{Digital Investigation}}  \bibinfo{volume}{22}
  (\bibinfo{year}{2017}), \bibinfo{pages}{S94--S105}.
\newblock


\bibitem[\protect\citeauthoryear{Grega, Matiola{\'n}ski, Guzik, and
  Leszczuk}{Grega et~al\mbox{.}}{2016}]%
        {grega2016automated}
\bibfield{author}{\bibinfo{person}{Micha{\l} Grega}, \bibinfo{person}{Andrzej
  Matiola{\'n}ski}, \bibinfo{person}{Piotr Guzik}, {and}
  \bibinfo{person}{Miko{\l}aj Leszczuk}.} \bibinfo{year}{2016}\natexlab{}.
\newblock \showarticletitle{Automated detection of firearms and knives in a
  CCTV image}.
\newblock \bibinfo{journal}{\emph{Sensors}} \bibinfo{volume}{16},
  \bibinfo{number}{1} (\bibinfo{year}{2016}), \bibinfo{pages}{47}.
\newblock


\bibitem[\protect\citeauthoryear{Hargreaves and Marshall}{Hargreaves and
  Marshall}{2019}]%
        {hargreaves2019synctriage}
\bibfield{author}{\bibinfo{person}{Christopher Hargreaves} {and}
  \bibinfo{person}{Angus Marshall}.} \bibinfo{year}{2019}\natexlab{}.
\newblock \showarticletitle{SyncTriage: Using synchronisation artefacts to
  optimise acquisition order}.
\newblock \bibinfo{journal}{\emph{Digital Investigation}}  \bibinfo{volume}{28}
  (\bibinfo{year}{2019}), \bibinfo{pages}{S134--S140}.
\newblock


\bibitem[\protect\citeauthoryear{Hargreaves and Patterson}{Hargreaves and
  Patterson}{2012}]%
        {hargreaves2012automated}
\bibfield{author}{\bibinfo{person}{Christopher Hargreaves} {and}
  \bibinfo{person}{Jonathan Patterson}.} \bibinfo{year}{2012}\natexlab{}.
\newblock \showarticletitle{An automated timeline reconstruction approach for
  digital forensic investigations}.
\newblock \bibinfo{journal}{\emph{Digital Investigation}}  \bibinfo{volume}{9}
  (\bibinfo{year}{2012}), \bibinfo{pages}{S69--S79}.
\newblock


\bibitem[\protect\citeauthoryear{Hospodar, Gierlichs, De~Mulder, Verbauwhede,
  and Vandewalle}{Hospodar et~al\mbox{.}}{2011}]%
        {hospodar2011machine}
\bibfield{author}{\bibinfo{person}{Gabriel Hospodar}, \bibinfo{person}{Benedikt
  Gierlichs}, \bibinfo{person}{Elke De~Mulder}, \bibinfo{person}{Ingrid
  Verbauwhede}, {and} \bibinfo{person}{Joos Vandewalle}.}
  \bibinfo{year}{2011}\natexlab{}.
\newblock \showarticletitle{Machine learning in side-channel analysis: a first
  study}.
\newblock \bibinfo{journal}{\emph{Journal of Cryptographic Engineering}}
  \bibinfo{volume}{1}, \bibinfo{number}{4} (\bibinfo{year}{2011}),
  \bibinfo{pages}{293}.
\newblock


\bibitem[\protect\citeauthoryear{Huang and Stokes}{Huang and Stokes}{2016}]%
        {huang2016mtnet}
\bibfield{author}{\bibinfo{person}{Wenyi Huang} {and} \bibinfo{person}{Jack~W
  Stokes}.} \bibinfo{year}{2016}\natexlab{}.
\newblock \showarticletitle{MtNet: a multi-task neural network for dynamic
  malware classification}. In \bibinfo{booktitle}{\emph{International
  conference on detection of intrusions and malware, and vulnerability
  assessment}}. Springer, \bibinfo{pages}{399--418}.
\newblock


\bibitem[\protect\citeauthoryear{Jasmine and Annadurai}{Jasmine and
  Annadurai}{2019}]%
        {JASMINE2019833}
\bibfield{author}{\bibinfo{person}{J. Jasmine} {and} \bibinfo{person}{S.
  Annadurai}.} \bibinfo{year}{2019}\natexlab{}.
\newblock \showarticletitle{Real time video image enhancement approach using
  particle swarm optimisation technique with adaptive cumulative distribution
  function based histogram equalisation}.
\newblock \bibinfo{journal}{\emph{Measurement}}  \bibinfo{volume}{145}
  (\bibinfo{year}{2019}), \bibinfo{pages}{833 -- 840}.
\newblock
\showISSN{0263-2241}
\urldef\tempurl%
\url{https://doi.org/10.1016/j.measurement.2018.12.105}
\showDOI{\tempurl}


\bibitem[\protect\citeauthoryear{Jeyaraman and Atallah}{Jeyaraman and
  Atallah}{2006}]%
        {jeyaraman2006empirical}
\bibfield{author}{\bibinfo{person}{Sundararaman Jeyaraman} {and}
  \bibinfo{person}{Mikhail~J Atallah}.} \bibinfo{year}{2006}\natexlab{}.
\newblock \showarticletitle{An empirical study of automatic event
  reconstruction systems}.
\newblock \bibinfo{journal}{\emph{digital investigation}}  \bibinfo{volume}{3}
  (\bibinfo{year}{2006}), \bibinfo{pages}{108--115}.
\newblock


\bibitem[\protect\citeauthoryear{Kalgutkar, Kaur, Gonzalez, Stakhanova, and
  Matyukhina}{Kalgutkar et~al\mbox{.}}{2019}]%
        {kalgutkar2019code}
\bibfield{author}{\bibinfo{person}{Vaibhavi Kalgutkar},
  \bibinfo{person}{Ratinder Kaur}, \bibinfo{person}{Hugo Gonzalez},
  \bibinfo{person}{Natalia Stakhanova}, {and} \bibinfo{person}{Alina
  Matyukhina}.} \bibinfo{year}{2019}\natexlab{}.
\newblock \showarticletitle{Code authorship attribution: Methods and
  challenges}.
\newblock \bibinfo{journal}{\emph{ACM Computing Surveys (CSUR)}}
  \bibinfo{volume}{52}, \bibinfo{number}{1} (\bibinfo{year}{2019}),
  \bibinfo{pages}{1--36}.
\newblock


\bibitem[\protect\citeauthoryear{Khan}{Khan}{2012}]%
        {khan2012performance}
\bibfield{author}{\bibinfo{person}{Muhammad Naeem~Ahmed Khan}.}
  \bibinfo{year}{2012}\natexlab{}.
\newblock \showarticletitle{Performance analysis of Bayesian networks and
  neural networks in classification of file system activities}.
\newblock \bibinfo{journal}{\emph{Computers \& Security}} \bibinfo{volume}{31},
  \bibinfo{number}{4} (\bibinfo{year}{2012}), \bibinfo{pages}{391--401}.
\newblock


\bibitem[\protect\citeauthoryear{Kharrazi, Sencar, and Memon}{Kharrazi
  et~al\mbox{.}}{2004}]%
        {kharrazi2004blind}
\bibfield{author}{\bibinfo{person}{Mehdi Kharrazi}, \bibinfo{person}{Husrev~T
  Sencar}, {and} \bibinfo{person}{Nasir Memon}.}
  \bibinfo{year}{2004}\natexlab{}.
\newblock \showarticletitle{Blind source camera identification}. In
  \bibinfo{booktitle}{\emph{Image Processing, 2004. ICIP'04. 2004 International
  Conference on}}, Vol.~\bibinfo{volume}{1}. IEEE, \bibinfo{pages}{709--712}.
\newblock


\bibitem[\protect\citeauthoryear{Kim, Lee, Choi, and Yoon}{Kim
  et~al\mbox{.}}{2016}]%
        {kim2016protecting}
\bibfield{author}{\bibinfo{person}{Taesung Kim}, \bibinfo{person}{Seungkwang
  Lee}, \bibinfo{person}{Dooho Choi}, {and} \bibinfo{person}{Hyunsoo Yoon}.}
  \bibinfo{year}{2016}\natexlab{}.
\newblock \showarticletitle{Protecting secret keys in networked devices with
  table encoding against power analysis attacks}.
\newblock \bibinfo{journal}{\emph{Journal of High Speed Networks}}
  \bibinfo{volume}{22}, \bibinfo{number}{4} (\bibinfo{year}{2016}),
  \bibinfo{pages}{293--307}.
\newblock


\bibitem[\protect\citeauthoryear{Kocher, Jaffe, and Jun}{Kocher
  et~al\mbox{.}}{1999}]%
        {kocher1999differential}
\bibfield{author}{\bibinfo{person}{Paul Kocher}, \bibinfo{person}{Joshua
  Jaffe}, {and} \bibinfo{person}{Benjamin Jun}.}
  \bibinfo{year}{1999}\natexlab{}.
\newblock \showarticletitle{Differential power analysis}. In
  \bibinfo{booktitle}{\emph{Advances in Cryptology (CRYPTO `99)}}. Springer,
  \bibinfo{pages}{789--789}.
\newblock


\bibitem[\protect\citeauthoryear{Kocher, Jaffe, Jun, and Rohatgi}{Kocher
  et~al\mbox{.}}{2011}]%
        {kocher2011introduction}
\bibfield{author}{\bibinfo{person}{Paul Kocher}, \bibinfo{person}{Joshua
  Jaffe}, \bibinfo{person}{Benjamin Jun}, {and} \bibinfo{person}{Pankaj
  Rohatgi}.} \bibinfo{year}{2011}\natexlab{}.
\newblock \showarticletitle{Introduction to differential power analysis}.
\newblock \bibinfo{journal}{\emph{Journal of Cryptographic Engineering}}
  \bibinfo{volume}{1}, \bibinfo{number}{1} (\bibinfo{year}{2011}),
  \bibinfo{pages}{5--27}.
\newblock


\bibitem[\protect\citeauthoryear{Kohn, Eloff, and Eloff}{Kohn
  et~al\mbox{.}}{2013}]%
        {kohn2013integrated}
\bibfield{author}{\bibinfo{person}{Michael~Donovan Kohn},
  \bibinfo{person}{Mariki~M Eloff}, {and} \bibinfo{person}{Jan~HP Eloff}.}
  \bibinfo{year}{2013}\natexlab{}.
\newblock \showarticletitle{Integrated digital forensic process model}.
\newblock \bibinfo{journal}{\emph{Computers \& Security}}  \bibinfo{volume}{38}
  (\bibinfo{year}{2013}), \bibinfo{pages}{103--115}.
\newblock


\bibitem[\protect\citeauthoryear{Kolosnjaji, Zarras, Webster, and
  Eckert}{Kolosnjaji et~al\mbox{.}}{2016}]%
        {kolosnjaji2016deep}
\bibfield{author}{\bibinfo{person}{Bojan Kolosnjaji},
  \bibinfo{person}{Apostolis Zarras}, \bibinfo{person}{George Webster}, {and}
  \bibinfo{person}{Claudia Eckert}.} \bibinfo{year}{2016}\natexlab{}.
\newblock \showarticletitle{Deep learning for classification of malware system
  call sequences}. In \bibinfo{booktitle}{\emph{Australasian Joint Conference
  on Artificial Intelligence}}. Springer, \bibinfo{pages}{137--149}.
\newblock


\bibitem[\protect\citeauthoryear{Laput, Yang, Xiao, Sample, and Harrison}{Laput
  et~al\mbox{.}}{2015}]%
        {laput2015sense}
\bibfield{author}{\bibinfo{person}{Gierad Laput}, \bibinfo{person}{Chouchang
  Yang}, \bibinfo{person}{Robert Xiao}, \bibinfo{person}{Alanson Sample}, {and}
  \bibinfo{person}{Chris Harrison}.} \bibinfo{year}{2015}\natexlab{}.
\newblock \showarticletitle{Em-sense: Touch recognition of uninstrumented,
  electrical and electromechanical objects}. In
  \bibinfo{booktitle}{\emph{Proceedings of the 28th Annual ACM Symposium on
  User Interface Software \& Technology}}. ACM, \bibinfo{pages}{157--166}.
\newblock


\bibitem[\protect\citeauthoryear{Lashkari, Draper-Gil, Mamun, and
  Ghorbani}{Lashkari et~al\mbox{.}}{2017}]%
        {Lashkari:2017}
\bibfield{author}{\bibinfo{person}{Arash~Habibi Lashkari},
  \bibinfo{person}{Gerard Draper-Gil}, \bibinfo{person}{Mohammad Saiful~Islam
  Mamun}, {and} \bibinfo{person}{Ali~A. Ghorbani}.}
  \bibinfo{year}{2017}\natexlab{}.
\newblock \showarticletitle{Characterization of Tor Traffic Using Time Based
  Features}. In \bibinfo{booktitle}{\emph{In the proceeding of the 3rd
  International Conference on Information System Security and Privacy,
  SCITEPRESS}} (Portugal).
\newblock


\bibitem[\protect\citeauthoryear{Le, Boydell, Mac~Namee, and Scanlon}{Le
  et~al\mbox{.}}{2018}]%
        {le2018deep}
\bibfield{author}{\bibinfo{person}{Quan Le}, \bibinfo{person}{Ois{\'\i}n
  Boydell}, \bibinfo{person}{Brian Mac~Namee}, {and} \bibinfo{person}{Mark
  Scanlon}.} \bibinfo{year}{2018}\natexlab{}.
\newblock \showarticletitle{Deep learning at the shallow end: Malware
  classification for non-domain experts}.
\newblock \bibinfo{journal}{\emph{Digital Investigation}}  \bibinfo{volume}{26}
  (\bibinfo{year}{2018}), \bibinfo{pages}{S118--S126}.
\newblock


\bibitem[\protect\citeauthoryear{Le, Cl{\'e}di{\`e}re, Serviere, and
  Lacoume}{Le et~al\mbox{.}}{2007}]%
        {le2007efficient}
\bibfield{author}{\bibinfo{person}{Thanh-Ha Le}, \bibinfo{person}{Jessy
  Cl{\'e}di{\`e}re}, \bibinfo{person}{Christine Serviere}, {and}
  \bibinfo{person}{Jean-Louis Lacoume}.} \bibinfo{year}{2007}\natexlab{}.
\newblock \showarticletitle{Efficient solution for misalignment of signal in
  side channel analysis}. In \bibinfo{booktitle}{\emph{2007 IEEE International
  Conference on Acoustics, Speech and Signal Processing-ICASSP'07}},
  Vol.~\bibinfo{volume}{2}. IEEE, \bibinfo{pages}{II--257}.
\newblock


\bibitem[\protect\citeauthoryear{LeCun, Bengio, and Hinton}{LeCun
  et~al\mbox{.}}{2015}]%
        {lecun2015deep}
\bibfield{author}{\bibinfo{person}{Yann LeCun}, \bibinfo{person}{Yoshua
  Bengio}, {and} \bibinfo{person}{Geoffrey Hinton}.}
  \bibinfo{year}{2015}\natexlab{}.
\newblock \showarticletitle{Deep learning}.
\newblock \bibinfo{journal}{\emph{nature}} \bibinfo{volume}{521},
  \bibinfo{number}{7553} (\bibinfo{year}{2015}), \bibinfo{pages}{436--444}.
\newblock


\bibitem[\protect\citeauthoryear{Lerman, Bontempi, and Markowitch}{Lerman
  et~al\mbox{.}}{2011}]%
        {lerman2011side}
\bibfield{author}{\bibinfo{person}{Liran Lerman}, \bibinfo{person}{Gianluca
  Bontempi}, {and} \bibinfo{person}{Olivier Markowitch}.}
  \bibinfo{year}{2011}\natexlab{}.
\newblock \showarticletitle{Side channel attack: an approach based on machine
  learning}. In \bibinfo{booktitle}{\emph{Proceedings of 2nd International
  Workshop on Constructive Side-Channel Analysis and Security Design
  (COSADE)}}. Schindler and Huss, \bibinfo{pages}{29--41}.
\newblock


\bibitem[\protect\citeauthoryear{Lessmann, Baesens, Seow, and Thomas}{Lessmann
  et~al\mbox{.}}{2015}]%
        {lessmann2015benchmarking}
\bibfield{author}{\bibinfo{person}{Stefan Lessmann}, \bibinfo{person}{Bart
  Baesens}, \bibinfo{person}{Hsin-Vonn Seow}, {and} \bibinfo{person}{Lyn~C
  Thomas}.} \bibinfo{year}{2015}\natexlab{}.
\newblock \showarticletitle{Benchmarking state-of-the-art classification
  algorithms for credit scoring: An update of research}.
\newblock \bibinfo{journal}{\emph{European Journal of Operational Research}}
  \bibinfo{volume}{247}, \bibinfo{number}{1} (\bibinfo{year}{2015}),
  \bibinfo{pages}{124--136}.
\newblock


\bibitem[\protect\citeauthoryear{Lillis, Becker, O'Sullivan, and
  Scanlon}{Lillis et~al\mbox{.}}{2016}]%
        {lillis2016challenges}
\bibfield{author}{\bibinfo{person}{David Lillis}, \bibinfo{person}{Brett
  Becker}, \bibinfo{person}{Tadhg O'Sullivan}, {and} \bibinfo{person}{Mark
  Scanlon}.} \bibinfo{year}{2016}\natexlab{}.
\newblock \showarticletitle{{Current Challenges and Future Research Areas for
  Digital Forensic Investigation}}. In \bibinfo{booktitle}{\emph{{The 11th
  ADFSL Conference on Digital Forensics, Security and Law (CDFSL 2016)}}}.
  \bibinfo{publisher}{ADFSL}, \bibinfo{address}{Daytona Beach, FL, USA},
  \bibinfo{pages}{9--20}.
\newblock


\bibitem[\protect\citeauthoryear{Liu, Xu, Zhang, and Wu}{Liu
  et~al\mbox{.}}{2018}]%
        {Liu:2018}
\bibfield{author}{\bibinfo{person}{Liqun Liu}, \bibinfo{person}{Bing Xu},
  \bibinfo{person}{Xiaoping Zhang}, {and} \bibinfo{person}{Xianjun Wu}.}
  \bibinfo{year}{2018}\natexlab{}.
\newblock \showarticletitle{An intrusion detection method for internet of
  things based on suppressed fuzzy clustering}.
\newblock \bibinfo{journal}{\emph{EURASIP Journal on Wireless Communications
  and Networking}} (\bibinfo{year}{2018}).
\newblock


\bibitem[\protect\citeauthoryear{Lukas, Fridrich, and Goljan}{Lukas
  et~al\mbox{.}}{2006}]%
        {lukas2006digital}
\bibfield{author}{\bibinfo{person}{Jan Lukas}, \bibinfo{person}{Jessica
  Fridrich}, {and} \bibinfo{person}{Miroslav Goljan}.}
  \bibinfo{year}{2006}\natexlab{}.
\newblock \showarticletitle{Digital camera identification from sensor pattern
  noise}.
\newblock \bibinfo{journal}{\emph{IEEE Transactions on Information Forensics
  and Security}} \bibinfo{volume}{1}, \bibinfo{number}{2}
  (\bibinfo{year}{2006}), \bibinfo{pages}{205--214}.
\newblock


\bibitem[\protect\citeauthoryear{MacEachern, Jindal-Snape, and
  Jackson}{MacEachern et~al\mbox{.}}{2011}]%
        {maceachern2011child}
\bibfield{author}{\bibinfo{person}{Alison~D MacEachern}, \bibinfo{person}{Divya
  Jindal-Snape}, {and} \bibinfo{person}{Sharon Jackson}.}
  \bibinfo{year}{2011}\natexlab{}.
\newblock \showarticletitle{Child abuse investigation: police officers and
  secondary traumatic stress}.
\newblock \bibinfo{journal}{\emph{International journal of occupational safety
  and ergonomics}} \bibinfo{volume}{17}, \bibinfo{number}{4}
  (\bibinfo{year}{2011}), \bibinfo{pages}{329--339}.
\newblock


\bibitem[\protect\citeauthoryear{Machado-Fern{\'a}ndez}{Machado-Fern{\'a}ndez}{2015}]%
        {machado2015software}
\bibfield{author}{\bibinfo{person}{Jos{\'e}~Ra{\'u}l Machado-Fern{\'a}ndez}.}
  \bibinfo{year}{2015}\natexlab{}.
\newblock \showarticletitle{Software defined radio: Basic principles and
  applications}.
\newblock \bibinfo{journal}{\emph{Revista Facultad de Ingenier{\'\i}a}}
  \bibinfo{volume}{24}, \bibinfo{number}{38} (\bibinfo{year}{2015}),
  \bibinfo{pages}{79--96}.
\newblock


\bibitem[\protect\citeauthoryear{Madry, Makelov, Schmidt, Tsipras, and
  Vladu}{Madry et~al\mbox{.}}{2017}]%
        {madry2017towards}
\bibfield{author}{\bibinfo{person}{Aleksander Madry},
  \bibinfo{person}{Aleksandar Makelov}, \bibinfo{person}{Ludwig Schmidt},
  \bibinfo{person}{Dimitris Tsipras}, {and} \bibinfo{person}{Adrian Vladu}.}
  \bibinfo{year}{2017}\natexlab{}.
\newblock \showarticletitle{Towards deep learning models resistant to
  adversarial attacks}.
\newblock \bibinfo{journal}{\emph{arXiv preprint arXiv:1706.06083}}
  (\bibinfo{year}{2017}).
\newblock


\bibitem[\protect\citeauthoryear{Maghrebi, Portigliatti, and Prouff}{Maghrebi
  et~al\mbox{.}}{2016}]%
        {maghrebi2016breaking}
\bibfield{author}{\bibinfo{person}{Houssem Maghrebi}, \bibinfo{person}{Thibault
  Portigliatti}, {and} \bibinfo{person}{Emmanuel Prouff}.}
  \bibinfo{year}{2016}\natexlab{}.
\newblock \showarticletitle{Breaking cryptographic implementations using deep
  learning techniques}. In \bibinfo{booktitle}{\emph{International Conference
  on Security, Privacy, and Applied Cryptography Engineering}}. Springer,
  \bibinfo{pages}{3--26}.
\newblock


\bibitem[\protect\citeauthoryear{Marrington, Baggili, Mohay, and
  Clark}{Marrington et~al\mbox{.}}{2011}]%
        {marrington2011cat}
\bibfield{author}{\bibinfo{person}{Andrew Marrington}, \bibinfo{person}{Ibrahim
  Baggili}, \bibinfo{person}{George Mohay}, {and} \bibinfo{person}{Andrew
  Clark}.} \bibinfo{year}{2011}\natexlab{}.
\newblock \showarticletitle{CAT Detect (Computer Activity Timeline Detection):
  A tool for detecting inconsistency in computer activity timelines}.
\newblock \bibinfo{journal}{\emph{digital investigation}}  \bibinfo{volume}{8}
  (\bibinfo{year}{2011}), \bibinfo{pages}{S52--S61}.
\newblock


\bibitem[\protect\citeauthoryear{Marturana, Me, Berte, and Tacconi}{Marturana
  et~al\mbox{.}}{2011}]%
        {marturana2011quantitative}
\bibfield{author}{\bibinfo{person}{Fabio Marturana}, \bibinfo{person}{Gianluigi
  Me}, \bibinfo{person}{Rosamaria Berte}, {and} \bibinfo{person}{Simone
  Tacconi}.} \bibinfo{year}{2011}\natexlab{}.
\newblock \showarticletitle{A quantitative approach to triaging in mobile
  forensics}. In \bibinfo{booktitle}{\emph{2011IEEE 10th International
  Conference on Trust, Security and Privacy in Computing and Communications}}.
  IEEE, \bibinfo{pages}{582--588}.
\newblock


\bibitem[\protect\citeauthoryear{Marturana and Tacconi}{Marturana and
  Tacconi}{2013}]%
        {marturana2013machine}
\bibfield{author}{\bibinfo{person}{Fabio Marturana} {and}
  \bibinfo{person}{Simone Tacconi}.} \bibinfo{year}{2013}\natexlab{}.
\newblock \showarticletitle{A Machine Learning-based Triage methodology for
  automated categorization of digital media}.
\newblock \bibinfo{journal}{\emph{Digital Investigation}} \bibinfo{volume}{10},
  \bibinfo{number}{2} (\bibinfo{year}{2013}), \bibinfo{pages}{193--204}.
\newblock


\bibitem[\protect\citeauthoryear{Masure, Dumas, and Prouff}{Masure
  et~al\mbox{.}}{2020}]%
        {masure2020comprehensive}
\bibfield{author}{\bibinfo{person}{Lo{\"\i}c Masure},
  \bibinfo{person}{C{\'e}cile Dumas}, {and} \bibinfo{person}{Emmanuel Prouff}.}
  \bibinfo{year}{2020}\natexlab{}.
\newblock \showarticletitle{A comprehensive study of deep learning for
  side-channel analysis}.
\newblock \bibinfo{journal}{\emph{IACR Transactions on Cryptographic Hardware
  and Embedded Systems}} (\bibinfo{year}{2020}), \bibinfo{pages}{348--375}.
\newblock


\bibitem[\protect\citeauthoryear{Meissner and Kassin}{Meissner and
  Kassin}{2002}]%
        {meissner2002he}
\bibfield{author}{\bibinfo{person}{Christian~A Meissner} {and}
  \bibinfo{person}{Saul~M Kassin}.} \bibinfo{year}{2002}\natexlab{}.
\newblock \showarticletitle{``He's guilty!'': Investigator bias in judgments of
  truth and deception}.
\newblock \bibinfo{journal}{\emph{Law and human behavior}}
  \bibinfo{volume}{26}, \bibinfo{number}{5} (\bibinfo{year}{2002}),
  \bibinfo{pages}{469--480}.
\newblock


\bibitem[\protect\citeauthoryear{Muhammad Naeem~Khan and Young}{Muhammad
  Naeem~Khan and Young}{2007}]%
        {khan2007framework}
\bibfield{author}{\bibinfo{person}{Chris R.~Chatwin Muhammad Naeem~Khan} {and}
  \bibinfo{person}{Rupert~CD Young}.} \bibinfo{year}{2007}\natexlab{}.
\newblock \showarticletitle{A framework for post-event timeline reconstruction
  using neural networks}.
\newblock \bibinfo{journal}{\emph{digital investigation}} \bibinfo{volume}{4},
  \bibinfo{number}{3-4} (\bibinfo{year}{2007}), \bibinfo{pages}{146--157}.
\newblock


\bibitem[\protect\citeauthoryear{Nath and Mehtre}{Nath and Mehtre}{2014}]%
        {nath2014static}
\bibfield{author}{\bibinfo{person}{Hiran~V Nath} {and} \bibinfo{person}{Babu~M
  Mehtre}.} \bibinfo{year}{2014}\natexlab{}.
\newblock \showarticletitle{Static malware analysis using machine learning
  methods}. In \bibinfo{booktitle}{\emph{International Conference on Security
  in Computer Networks and Distributed Systems}}. Springer,
  \bibinfo{pages}{440--450}.
\newblock


\bibitem[\protect\citeauthoryear{Naviq, Azwar, Ali, and Rehman}{Naviq
  et~al\mbox{.}}{2018}]%
        {Murtaz:2018}
\bibfield{author}{\bibinfo{person}{Mohammed Murtaz~Amir Naviq},
  \bibinfo{person}{Hassan Azwar}, \bibinfo{person}{Syed~Baqir Ali}, {and}
  \bibinfo{person}{Saad Rehman}.} \bibinfo{year}{2018}\natexlab{}.
\newblock \showarticletitle{A framework for Android Malware detection and
  classification}. In \bibinfo{booktitle}{\emph{2018 IEEE 5th International
  Conference on Engineering Technologies and Applied Sciences (ICETAS)}}.
  \bibinfo{pages}{1--5}.
\newblock


\bibitem[\protect\citeauthoryear{Nazari, Sehatbakhsh, Alam, Zajic, and
  Prvulovic}{Nazari et~al\mbox{.}}{2017}]%
        {nazari2017eddie}
\bibfield{author}{\bibinfo{person}{Alireza Nazari}, \bibinfo{person}{Nader
  Sehatbakhsh}, \bibinfo{person}{Monjur Alam}, \bibinfo{person}{Alenka Zajic},
  {and} \bibinfo{person}{Milos Prvulovic}.} \bibinfo{year}{2017}\natexlab{}.
\newblock \showarticletitle{{EDDIE: EM-Based Detection of Deviations in Program
  Execution}}. In \bibinfo{booktitle}{\emph{Proceedings of the 44th Annual
  International Symposium on Computer Architecture}}. ACM,
  \bibinfo{pages}{333--346}.
\newblock


\bibitem[\protect\citeauthoryear{Nian, Li, Wang, Xu, and Wu}{Nian
  et~al\mbox{.}}{2016}]%
        {NIAN2016283}
\bibfield{author}{\bibinfo{person}{Fudong Nian}, \bibinfo{person}{Teng Li},
  \bibinfo{person}{Yan Wang}, \bibinfo{person}{Mingliang Xu}, {and}
  \bibinfo{person}{Jun Wu}.} \bibinfo{year}{2016}\natexlab{}.
\newblock \showarticletitle{Pornographic image detection utilizing deep
  convolutional neural networks}.
\newblock \bibinfo{journal}{\emph{Neurocomputing}}  \bibinfo{volume}{210}
  (\bibinfo{year}{2016}), \bibinfo{pages}{283 -- 293}.
\newblock
\showISSN{0925-2312}
\urldef\tempurl%
\url{https://doi.org/10.1016/j.neucom.2015.09.135}
\showDOI{\tempurl}
\newblock
\shownote{SI:Behavior Analysis In SN.}


\bibitem[\protect\citeauthoryear{Nour~Moustafa and Slay}{Nour~Moustafa and
  Slay}{2019}]%
        {Moustafa:2019}
\bibfield{author}{\bibinfo{person}{Jiankun~Hu Nour~Moustafa} {and}
  \bibinfo{person}{Jill Slay}.} \bibinfo{year}{2019}\natexlab{}.
\newblock \showarticletitle{A holistic review of Network Anomaly Detection
  Systems: {A} comprehensive survey}.
\newblock \bibinfo{journal}{\emph{J. Network and Computer Applications}}
  \bibinfo{volume}{128} (\bibinfo{year}{2019}), \bibinfo{pages}{33--55}.
\newblock


\bibitem[\protect\citeauthoryear{Oh, Schiele, and Fritz}{Oh
  et~al\mbox{.}}{2019}]%
        {oh2019towards}
\bibfield{author}{\bibinfo{person}{Seong~Joon Oh}, \bibinfo{person}{Bernt
  Schiele}, {and} \bibinfo{person}{Mario Fritz}.}
  \bibinfo{year}{2019}\natexlab{}.
\newblock \showarticletitle{Towards reverse-engineering black-box neural
  networks}.
\newblock In \bibinfo{booktitle}{\emph{Explainable AI: Interpreting, Explaining
  and Visualizing Deep Learning}}. \bibinfo{publisher}{Springer},
  \bibinfo{pages}{121--144}.
\newblock


\bibitem[\protect\citeauthoryear{Pour, Bou-Harb, Varma, Neshenko, Pados, and
  Choo}{Pour et~al\mbox{.}}{2019}]%
        {SAFAEIPOUR:2019}
\bibfield{author}{\bibinfo{person}{Morteza~Safaei Pour}, \bibinfo{person}{Elias
  Bou-Harb}, \bibinfo{person}{Kavita Varma}, \bibinfo{person}{Nataliia
  Neshenko}, \bibinfo{person}{Dimitris~A. Pados}, {and}
  \bibinfo{person}{Kim-Kwang~Raymond Choo}.} \bibinfo{year}{2019}\natexlab{}.
\newblock \showarticletitle{Comprehending the IoT cyber threat landscape: A
  data dimensionality reduction technique to infer and characterize
  Internet-scale IoT probing campaigns}.
\newblock \bibinfo{journal}{\emph{Digital Investigation}}  \bibinfo{volume}{28}
  (\bibinfo{year}{2019}), \bibinfo{pages}{S40 -- S49}.
\newblock
\showISSN{1742-2876}
\urldef\tempurl%
\url{https://doi.org/10.1016/j.diin.2019.01.014}
\showDOI{\tempurl}


\bibitem[\protect\citeauthoryear{Rao and Ni}{Rao and Ni}{2016}]%
        {rao2016deep}
\bibfield{author}{\bibinfo{person}{Yuan Rao} {and} \bibinfo{person}{Jiangqun
  Ni}.} \bibinfo{year}{2016}\natexlab{}.
\newblock \showarticletitle{A deep learning approach to detection of splicing
  and copy-move forgeries in images}. In \bibinfo{booktitle}{\emph{Information
  Forensics and Security (WIFS), 2016 IEEE International Workshop on}}. IEEE,
  \bibinfo{pages}{1--6}.
\newblock


\bibitem[\protect\citeauthoryear{Rene and Abdullah}{Rene and Abdullah}{2017}]%
        {Rene:2017}
\bibfield{author}{\bibinfo{person}{Chiadighikaobi~Ikenna Rene} {and}
  \bibinfo{person}{Johari Abdullah}.} \bibinfo{year}{2017}\natexlab{}.
\newblock \showarticletitle{Malicious Code Intrusion Detection using Machine
  Learning And Indicators of Compromise}.
\newblock \bibinfo{journal}{\emph{International Journal of Computer Science and
  Information Security (IJCSIS)}} \bibinfo{volume}{15}, \bibinfo{number}{9}
  (\bibinfo{date}{September} \bibinfo{year}{2017}).
\newblock


\bibitem[\protect\citeauthoryear{Rivest}{Rivest}{1991}]%
        {rivest1991cryptography}
\bibfield{author}{\bibinfo{person}{Ronald~L Rivest}.}
  \bibinfo{year}{1991}\natexlab{}.
\newblock \showarticletitle{Cryptography and machine learning}. In
  \bibinfo{booktitle}{\emph{International Conference on the Theory and
  Application of Cryptology}}. Springer, \bibinfo{pages}{427--439}.
\newblock


\bibitem[\protect\citeauthoryear{Rocha, Scheirer, Forstall, Cavalcante,
  Theophilo, Shen, Carvalho, and Stamatatos}{Rocha et~al\mbox{.}}{2016}]%
        {rocha2016authorship}
\bibfield{author}{\bibinfo{person}{Anderson Rocha}, \bibinfo{person}{Walter~J
  Scheirer}, \bibinfo{person}{Christopher~W Forstall}, \bibinfo{person}{Thiago
  Cavalcante}, \bibinfo{person}{Antonio Theophilo}, \bibinfo{person}{Bingyu
  Shen}, \bibinfo{person}{Ariadne~RB Carvalho}, {and}
  \bibinfo{person}{Efstathios Stamatatos}.} \bibinfo{year}{2016}\natexlab{}.
\newblock \showarticletitle{Authorship attribution for social media forensics}.
\newblock \bibinfo{journal}{\emph{IEEE Transactions on Information Forensics
  and Security}} \bibinfo{volume}{12}, \bibinfo{number}{1}
  (\bibinfo{year}{2016}), \bibinfo{pages}{5--33}.
\newblock


\bibitem[\protect\citeauthoryear{Rogers, Goldman, Mislan, Wedge, and
  Debrota}{Rogers et~al\mbox{.}}{2006}]%
        {rogers2006computer}
\bibfield{author}{\bibinfo{person}{Marcus~K Rogers}, \bibinfo{person}{James
  Goldman}, \bibinfo{person}{Rick Mislan}, \bibinfo{person}{Timothy Wedge},
  {and} \bibinfo{person}{Steve Debrota}.} \bibinfo{year}{2006}\natexlab{}.
\newblock \showarticletitle{Computer forensics field triage process model}.
\newblock \bibinfo{journal}{\emph{Journal of Digital Forensics, Security and
  Law}} \bibinfo{volume}{1}, \bibinfo{number}{2} (\bibinfo{year}{2006}),
  \bibinfo{pages}{2}.
\newblock


\bibitem[\protect\citeauthoryear{Ruder}{Ruder}{2016}]%
        {ruder2016overview}
\bibfield{author}{\bibinfo{person}{Sebastian Ruder}.}
  \bibinfo{year}{2016}\natexlab{}.
\newblock \showarticletitle{An overview of gradient descent optimization
  algorithms}.
\newblock \bibinfo{journal}{\emph{arXiv preprint arXiv:1609.04747}}
  (\bibinfo{year}{2016}).
\newblock


\bibitem[\protect\citeauthoryear{Sanchez, Grajeda, Baggili, and Hall}{Sanchez
  et~al\mbox{.}}{2019}]%
        {sanchez2019practitioner}
\bibfield{author}{\bibinfo{person}{Laura Sanchez}, \bibinfo{person}{Cinthya
  Grajeda}, \bibinfo{person}{Ibrahim Baggili}, {and} \bibinfo{person}{Cory
  Hall}.} \bibinfo{year}{2019}\natexlab{}.
\newblock \showarticletitle{A Practitioner Survey Exploring the Value of
  Forensic Tools, AI, Filtering, \& Safer Presentation for Investigating Child
  Sexual Abuse Material (CSAM)}.
\newblock \bibinfo{journal}{\emph{Digital Investigation}}  \bibinfo{volume}{29}
  (\bibinfo{year}{2019}), \bibinfo{pages}{S124--S142}.
\newblock


\bibitem[\protect\citeauthoryear{Saputra, Vijaykrishnan, Kandemir, Irwin,
  Brooks, Kim, and Zhang}{Saputra et~al\mbox{.}}{2003}]%
        {saputra2003masking}
\bibfield{author}{\bibinfo{person}{Hendra Saputra}, \bibinfo{person}{Narayanan
  Vijaykrishnan}, \bibinfo{person}{M Kandemir}, \bibinfo{person}{Mary~Jane
  Irwin}, \bibinfo{person}{R Brooks}, \bibinfo{person}{Soontae Kim}, {and}
  \bibinfo{person}{Wei Zhang}.} \bibinfo{year}{2003}\natexlab{}.
\newblock \showarticletitle{Masking the energy behavior of DES encryption}. In
  \bibinfo{booktitle}{\emph{Proceedings of the conference on Design, Automation
  and Test in Europe-Volume 1}}. IEEE Computer Society, \bibinfo{pages}{10084}.
\newblock


\bibitem[\protect\citeauthoryear{Saud Mohammed~Othman and Zahary}{Saud
  Mohammed~Othman and Zahary}{2018}]%
        {Othman:2018}
\bibfield{author}{\bibinfo{person}{Fadl Mutaher Ba-Alwi Saud Mohammed~Othman,
  Nabeel T~Alsohybe} {and} \bibinfo{person}{Ammar~Thabit Zahary}.}
  \bibinfo{year}{2018}\natexlab{}.
\newblock \showarticletitle{Survey on Intrusion Detection System}.
\newblock \bibinfo{journal}{\emph{International Journal of Cyber-Security and
  Digital Forensics (IJCSDF)}} (\bibinfo{date}{December} \bibinfo{year}{2018}).
\newblock
\showISSN{2305-001}


\bibitem[\protect\citeauthoryear{Sayakkara, Le-Khac, and Scanlon}{Sayakkara
  et~al\mbox{.}}{2019}]%
        {sayakkara2019survey}
\bibfield{author}{\bibinfo{person}{Asanka Sayakkara}, \bibinfo{person}{Nhien-An
  Le-Khac}, {and} \bibinfo{person}{Mark Scanlon}.}
  \bibinfo{year}{2019}\natexlab{}.
\newblock \showarticletitle{A survey of electromagnetic side-channel attacks
  and discussion on their case-progressing potential for digital forensics}.
\newblock \bibinfo{journal}{\emph{Digital Investigation}}
  (\bibinfo{year}{2019}).
\newblock


\bibitem[\protect\citeauthoryear{Scanlon}{Scanlon}{2016}]%
        {scanlon2016battling}
\bibfield{author}{\bibinfo{person}{Mark Scanlon}.}
  \bibinfo{year}{2016}\natexlab{}.
\newblock \showarticletitle{Battling the digital forensic backlog through data
  deduplication}. In \bibinfo{booktitle}{\emph{2016 Sixth International
  Conference on Innovative Computing Technology (INTECH)}}. IEEE,
  \bibinfo{pages}{10--14}.
\newblock


\bibitem[\protect\citeauthoryear{Schneider and Breitinger}{Schneider and
  Breitinger}{2020}]%
        {schneiderai}
\bibfield{author}{\bibinfo{person}{Johannes Schneider} {and}
  \bibinfo{person}{Frank Breitinger}.} \bibinfo{year}{2020}\natexlab{}.
\newblock \showarticletitle{AI Forensics: Did the Artificial Intelligence
  System Do It? Why?}
\newblock  (\bibinfo{year}{2020}).
\newblock


\bibitem[\protect\citeauthoryear{Sencar and Memon}{Sencar and Memon}{2009}]%
        {sencar2009overview}
\bibfield{author}{\bibinfo{person}{Husrev~T Sencar} {and}
  \bibinfo{person}{Nasir Memon}.} \bibinfo{year}{2009}\natexlab{}.
\newblock \showarticletitle{Overview of state-of-the-art in digital image
  forensics}.
\newblock In \bibinfo{booktitle}{\emph{Algorithms, Architectures and
  Information Systems Security}}. \bibinfo{publisher}{World Scientific},
  \bibinfo{pages}{325--347}.
\newblock


\bibitem[\protect\citeauthoryear{Shalev-Shwartz and Ben-David}{Shalev-Shwartz
  and Ben-David}{2014}]%
        {shalev2014understanding}
\bibfield{author}{\bibinfo{person}{Shai Shalev-Shwartz} {and}
  \bibinfo{person}{Shai Ben-David}.} \bibinfo{year}{2014}\natexlab{}.
\newblock \bibinfo{booktitle}{\emph{Understanding machine learning: From theory
  to algorithms}}.
\newblock \bibinfo{publisher}{Cambridge university press}.
\newblock


\bibitem[\protect\citeauthoryear{Shankar, Narumanchi, Ananya, Kompalli, and
  Chaudhury}{Shankar et~al\mbox{.}}{2017}]%
        {shankar2017deep}
\bibfield{author}{\bibinfo{person}{Devashish Shankar}, \bibinfo{person}{Sujay
  Narumanchi}, \bibinfo{person}{HA Ananya}, \bibinfo{person}{Pramod Kompalli},
  {and} \bibinfo{person}{Krishnendu Chaudhury}.}
  \bibinfo{year}{2017}\natexlab{}.
\newblock \showarticletitle{Deep learning based large scale visual
  recommendation and search for e-commerce}.
\newblock \bibinfo{journal}{\emph{arXiv preprint arXiv:1703.02344}}
  (\bibinfo{year}{2017}).
\newblock


\bibitem[\protect\citeauthoryear{Sharafaldin, Lashkari, and
  Ghorbani}{Sharafaldin et~al\mbox{.}}{2018}]%
        {Sharafaldin:2018}
\bibfield{author}{\bibinfo{person}{Iman Sharafaldin},
  \bibinfo{person}{Arash~Habibi Lashkari}, {and} \bibinfo{person}{Ali~A.
  Ghorbani}.} \bibinfo{year}{2018}\natexlab{}.
\newblock \showarticletitle{Toward Generating a New Intrusion Detection Dataset
  and Intrusion Traffic Characterization}. In \bibinfo{booktitle}{\emph{4th
  International Conference on Information Systems Security and Privacy
  (ICISSP)}} (Portugal).
\newblock


\bibitem[\protect\citeauthoryear{Sharafaldin, Lashkari, and
  Ghorbani}{Sharafaldin et~al\mbox{.}}{2019}]%
        {Sharafaldin:2019}
\bibfield{author}{\bibinfo{person}{Iman Sharafaldin},
  \bibinfo{person}{Arash~Habibi Lashkari}, {and} \bibinfo{person}{Ali~A.
  Ghorbani}.} \bibinfo{year}{2019}\natexlab{}.
\newblock \showarticletitle{Developing Realistic Distributed Denial of Service
  (DDoS) Attack Dataset and Taxonomy}. In \bibinfo{booktitle}{\emph{IEEE 53rd
  International Carnahan Conference on Security Technology}} (India).
\newblock


\bibitem[\protect\citeauthoryear{Singh, Singh, and Kaur}{Singh
  et~al\mbox{.}}{2019}]%
        {SINGH:2019}
\bibfield{author}{\bibinfo{person}{Manmeet Singh}, \bibinfo{person}{Maninder
  Singh}, {and} \bibinfo{person}{Sanmeet Kaur}.}
  \bibinfo{year}{2019}\natexlab{}.
\newblock \showarticletitle{Detecting bot-infected machines using DNS
  fingerprinting}.
\newblock \bibinfo{journal}{\emph{Digital Investigation}}  \bibinfo{volume}{28}
  (\bibinfo{year}{2019}), \bibinfo{pages}{14 -- 33}.
\newblock
\showISSN{1742-2876}
\urldef\tempurl%
\url{https://doi.org/10.1016/j.diin.2018.12.005}
\showDOI{\tempurl}


\bibitem[\protect\citeauthoryear{Spreitzer, Moonsamy, Korak, and
  Mangard}{Spreitzer et~al\mbox{.}}{2018}]%
        {spreitzer2018systematic}
\bibfield{author}{\bibinfo{person}{Raphael Spreitzer},
  \bibinfo{person}{Veelasha Moonsamy}, \bibinfo{person}{Thomas Korak}, {and}
  \bibinfo{person}{Stefan Mangard}.} \bibinfo{year}{2018}\natexlab{}.
\newblock \showarticletitle{Systematic classification of side-channel attacks:
  a case study for mobile devices}.
\newblock \bibinfo{journal}{\emph{IEEE Communications Surveys \& Tutorials}}
  \bibinfo{volume}{20}, \bibinfo{number}{1} (\bibinfo{year}{2018}),
  \bibinfo{pages}{465--488}.
\newblock


\bibitem[\protect\citeauthoryear{Stone and Stone}{Stone and Stone}{2016}]%
        {stone2016comparison}
\bibfield{author}{\bibinfo{person}{Barron Stone} {and} \bibinfo{person}{Samuel
  Stone}.} \bibinfo{year}{2016}\natexlab{}.
\newblock \showarticletitle{Comparison of Radio Frequency Based Techniques for
  Device Discrimination and Operation Identification}. In
  \bibinfo{booktitle}{\emph{11th International Conference on Cyber Warfare and
  Security: ICCWS2016}}. Academic Conferences and Publishing Limited,
  \bibinfo{pages}{475}.
\newblock


\bibitem[\protect\citeauthoryear{Studiawan, Sohel, and Payne}{Studiawan
  et~al\mbox{.}}{2020}]%
        {studiawan2020sentiment}
\bibfield{author}{\bibinfo{person}{Hudan Studiawan}, \bibinfo{person}{Ferdous
  Sohel}, {and} \bibinfo{person}{Christian Payne}.}
  \bibinfo{year}{2020}\natexlab{}.
\newblock \showarticletitle{Sentiment Analysis in a Forensic Timeline with Deep
  Learning}.
\newblock \bibinfo{journal}{\emph{IEEE Access}} (\bibinfo{year}{2020}).
\newblock


\bibitem[\protect\citeauthoryear{Stylianou, Schreier, Souvenir, and
  Pless}{Stylianou et~al\mbox{.}}{2017}]%
        {stylianou2017traffickcam}
\bibfield{author}{\bibinfo{person}{Abby Stylianou}, \bibinfo{person}{Jessica
  Schreier}, \bibinfo{person}{Richard Souvenir}, {and} \bibinfo{person}{Robert
  Pless}.} \bibinfo{year}{2017}\natexlab{}.
\newblock \showarticletitle{Traffickcam: Crowdsourced and computer vision based
  approaches to fighting sex trafficking}. In \bibinfo{booktitle}{\emph{2017
  IEEE Applied Imagery Pattern Recognition Workshop (AIPR)}}. IEEE,
  \bibinfo{pages}{1--8}.
\newblock


\bibitem[\protect\citeauthoryear{Taheri, Kadir, and Lashkari}{Taheri
  et~al\mbox{.}}{2019}]%
        {Taheri:2019}
\bibfield{author}{\bibinfo{person}{Laya Taheri}, \bibinfo{person}{Andi
  Fitriah~Abdul Kadir}, {and} \bibinfo{person}{Arash~Habibi Lashkari}.}
  \bibinfo{year}{2019}\natexlab{}.
\newblock \showarticletitle{Extensible Android Malware Detection and Family
  Classification Using Network-Flows and API-Calls}. In
  \bibinfo{booktitle}{\emph{2019 International Carnahan Conference on Security
  Technology (ICCST)}}. \bibinfo{pages}{1--8}.
\newblock


\bibitem[\protect\citeauthoryear{Tian and Huss}{Tian and Huss}{2012}]%
        {tian2012clock}
\bibfield{author}{\bibinfo{person}{Qizhi Tian} {and} \bibinfo{person}{Sorin~A
  Huss}.} \bibinfo{year}{2012}\natexlab{}.
\newblock \showarticletitle{On clock frequency effects in side channel attacks
  of symmetric block ciphers}. In \bibinfo{booktitle}{\emph{2012 5th
  International Conference on New Technologies, Mobility and Security (NTMS)}}.
  IEEE, \bibinfo{pages}{1--5}.
\newblock


\bibitem[\protect\citeauthoryear{Tsai, Lai, and Liu}{Tsai
  et~al\mbox{.}}{2007}]%
        {Tsai2007}
\bibfield{author}{\bibinfo{person}{Min~Jen Tsai}, \bibinfo{person}{Cheng~Liang
  Lai}, {and} \bibinfo{person}{Jung Liu}.} \bibinfo{year}{2007}\natexlab{}.
\newblock \showarticletitle{{Camera/mobile phone source identification for
  digital forensics}}.
\newblock \bibinfo{journal}{\emph{ICASSP, IEEE International Conference on
  Acoustics, Speech and Signal Processing - Proceedings}}  \bibinfo{volume}{2}
  (\bibinfo{year}{2007}), \bibinfo{pages}{221--224}.
\newblock
\showISBNx{1424407281}
\showISSN{15206149}
\urldef\tempurl%
\url{https://doi.org/10.1109/ICASSP.2007.366212}
\showDOI{\tempurl}


\bibitem[\protect\citeauthoryear{Turnbull and Randhawa}{Turnbull and
  Randhawa}{2015}]%
        {turnbull2015automated}
\bibfield{author}{\bibinfo{person}{Benjamin Turnbull} {and}
  \bibinfo{person}{Suneel Randhawa}.} \bibinfo{year}{2015}\natexlab{}.
\newblock \showarticletitle{Automated event and social network extraction from
  digital evidence sources with ontological mapping}.
\newblock \bibinfo{journal}{\emph{Digital Investigation}}  \bibinfo{volume}{13}
  (\bibinfo{year}{2015}), \bibinfo{pages}{94--106}.
\newblock


\bibitem[\protect\citeauthoryear{Ustebay, Turgutand, and Aydin}{Ustebay
  et~al\mbox{.}}{2018}]%
        {Ustebay:2018}
\bibfield{author}{\bibinfo{person}{Serpil Ustebay}, \bibinfo{person}{Zeynep
  Turgutand}, {and} \bibinfo{person}{Muhammed~Ali Aydin}.}
  \bibinfo{year}{2018}\natexlab{}.
\newblock \showarticletitle{Intrusion Detection System with Recursive Feature
  Elimination by Using Random Forest and Deep Learning Classifier}. In
  \bibinfo{booktitle}{\emph{2018 International Congress on Big Data, Deep
  Learning and Fighting Cyber Terrorism (IBIGDELFT)}}. \bibinfo{pages}{71--76}.
\newblock
\urldef\tempurl%
\url{https://doi.org/10.1109/IBIGDELFT.2018.8625318}
\showDOI{\tempurl}


\bibitem[\protect\citeauthoryear{Veale, Binns, and Edwards}{Veale
  et~al\mbox{.}}{2018}]%
        {veale2018algorithms}
\bibfield{author}{\bibinfo{person}{Michael Veale}, \bibinfo{person}{Reuben
  Binns}, {and} \bibinfo{person}{Lilian Edwards}.}
  \bibinfo{year}{2018}\natexlab{}.
\newblock \showarticletitle{Algorithms that remember: model inversion attacks
  and data protection law}.
\newblock \bibinfo{journal}{\emph{Philosophical Transactions of the Royal
  Society A: Mathematical, Physical and Engineering Sciences}}
  \bibinfo{volume}{376}, \bibinfo{number}{2133} (\bibinfo{year}{2018}),
  \bibinfo{pages}{20180083}.
\newblock


\bibitem[\protect\citeauthoryear{Vesel\'{y} and \v{Z}\'{a}dn\'{i}k}{Vesel\'{y}
  and \v{Z}\'{a}dn\'{i}k}{2019}]%
        {FITPUB12057}
\bibfield{author}{\bibinfo{person}{Vladim\'{i}r Vesel\'{y}} {and}
  \bibinfo{person}{Martin \v{Z}\'{a}dn\'{i}k}.}
  \bibinfo{year}{2019}\natexlab{}.
\newblock \showarticletitle{How to detect cryptocurrency miners? By traffic
  forensics!}
\newblock \bibinfo{journal}{\emph{Digital Investigation}} \bibinfo{volume}{31},
  \bibinfo{number}{31} (\bibinfo{year}{2019}), \bibinfo{pages}{1--25}.
\newblock
\showISSN{1742-2876}
\urldef\tempurl%
\url{https://doi.org/10.1016/j.diin.2019.08.002}
\showDOI{\tempurl}


\bibitem[\protect\citeauthoryear{Vinayakumar, Alazab, Soman, Poornachandran,
  Al-Nemrat, and Venkatraman}{Vinayakumar et~al\mbox{.}}{2019}]%
        {Vinayakumar:2019}
\bibfield{author}{\bibinfo{person}{R. Vinayakumar}, \bibinfo{person}{Mamoun
  Alazab}, \bibinfo{person}{K.~P. Soman}, \bibinfo{person}{Prabaharan
  Poornachandran}, \bibinfo{person}{Ameer Al-Nemrat}, {and}
  \bibinfo{person}{Sitalakshmi Venkatraman}.} \bibinfo{year}{2019}\natexlab{}.
\newblock \showarticletitle{Deep Learning Approach for Intelligent Intrusion
  Detection System}.
\newblock \bibinfo{journal}{\emph{IEEE Access}}  \bibinfo{volume}{7}
  (\bibinfo{year}{2019}), \bibinfo{pages}{41525--41550}.
\newblock
\showISSN{2169-3536}
\urldef\tempurl%
\url{https://doi.org/10.1109/ACCESS.2019.2895334}
\showDOI{\tempurl}


\bibitem[\protect\citeauthoryear{Vincze}{Vincze}{2016}]%
        {vincze2016challenges}
\bibfield{author}{\bibinfo{person}{Eva~A Vincze}.}
  \bibinfo{year}{2016}\natexlab{}.
\newblock \showarticletitle{Challenges in digital forensics}.
\newblock \bibinfo{journal}{\emph{Police Practice and Research}}
  \bibinfo{volume}{17}, \bibinfo{number}{2} (\bibinfo{year}{2016}),
  \bibinfo{pages}{183--194}.
\newblock


\bibitem[\protect\citeauthoryear{Vulinovi{\'c}, Ivkovi{\'c}, Petrovi{\'c},
  Skra{\v{c}}i{\'c}, and Pale}{Vulinovi{\'c} et~al\mbox{.}}{2019}]%
        {vulinovic2019neural}
\bibfield{author}{\bibinfo{person}{Kristijan Vulinovi{\'c}},
  \bibinfo{person}{Lucija Ivkovi{\'c}}, \bibinfo{person}{Juraj Petrovi{\'c}},
  \bibinfo{person}{Kristian Skra{\v{c}}i{\'c}}, {and} \bibinfo{person}{Predrag
  Pale}.} \bibinfo{year}{2019}\natexlab{}.
\newblock \showarticletitle{Neural Networks for File Fragment Classification}.
  In \bibinfo{booktitle}{\emph{2019 42nd International Convention on
  Information and Communication Technology, Electronics and Microelectronics
  (MIPRO)}}. IEEE, \bibinfo{pages}{1194--1198}.
\newblock


\bibitem[\protect\citeauthoryear{Wang and Wang}{Wang and Wang}{2014}]%
        {wang2014improving}
\bibfield{author}{\bibinfo{person}{Xinxi Wang} {and} \bibinfo{person}{Ye
  Wang}.} \bibinfo{year}{2014}\natexlab{}.
\newblock \showarticletitle{Improving content-based and hybrid music
  recommendation using deep learning}. In \bibinfo{booktitle}{\emph{Proceedings
  of the 22nd ACM international conference on Multimedia}}.
  \bibinfo{pages}{627--636}.
\newblock


\bibitem[\protect\citeauthoryear{Wang, Zhou, Harer, Brown, Qiu, Dou, Wang,
  Hinton, Gonzalez, and Chin}{Wang et~al\mbox{.}}{2018}]%
        {wang2018deep}
\bibfield{author}{\bibinfo{person}{Xiao Wang}, \bibinfo{person}{Quan Zhou},
  \bibinfo{person}{Jacob Harer}, \bibinfo{person}{Gavin Brown},
  \bibinfo{person}{Shangran Qiu}, \bibinfo{person}{Zhi Dou},
  \bibinfo{person}{John Wang}, \bibinfo{person}{Alan Hinton},
  \bibinfo{person}{Carlos~Aguayo Gonzalez}, {and} \bibinfo{person}{Peter
  Chin}.} \bibinfo{year}{2018}\natexlab{}.
\newblock \showarticletitle{Deep learning-based classification and anomaly
  detection of side-channel signals}. In \bibinfo{booktitle}{\emph{Cyber
  Sensing 2018}}, Vol.~\bibinfo{volume}{10630}. International Society for
  Optics and Photonics, \bibinfo{pages}{1063006}.
\newblock


\bibitem[\protect\citeauthoryear{Wolak and Mitchell}{Wolak and
  Mitchell}{2009}]%
        {wolak2009work}
\bibfield{author}{\bibinfo{person}{Janis Wolak} {and}
  \bibinfo{person}{Kimberly~J Mitchell}.} \bibinfo{year}{2009}\natexlab{}.
\newblock \showarticletitle{Work exposure to child pornography in ICAC task
  forces and affiliates}.
\newblock \bibinfo{journal}{\emph{Retrieved from Crimes against Children
  Research Center: http://www. unh. edu/ccrc/pdf/Law\% 20Enforcement\% 20Work\%
  20Exposure\% 20to\% 20CP. pdf}} (\bibinfo{year}{2009}).
\newblock


\bibitem[\protect\citeauthoryear{Xiao, Li, and Xu}{Xiao et~al\mbox{.}}{2019}]%
        {xiao2019video}
\bibfield{author}{\bibinfo{person}{Jianyu Xiao}, \bibinfo{person}{Shancang Li},
  {and} \bibinfo{person}{Qingliang Xu}.} \bibinfo{year}{2019}\natexlab{}.
\newblock \showarticletitle{Video-based evidence analysis and extraction in
  digital forensic investigation}.
\newblock \bibinfo{journal}{\emph{IEEE Access}}  \bibinfo{volume}{7}
  (\bibinfo{year}{2019}), \bibinfo{pages}{55432--55442}.
\newblock


\bibitem[\protect\citeauthoryear{Yang and Luo}{Yang and Luo}{2017}]%
        {10.1145/3011871}
\bibfield{author}{\bibinfo{person}{Xitong Yang} {and} \bibinfo{person}{Jiebo
  Luo}.} \bibinfo{year}{2017}\natexlab{}.
\newblock \showarticletitle{Tracking Illicit Drug Dealing and Abuse on
  Instagram Using Multimodal Analysis}.
\newblock \bibinfo{journal}{\emph{ACM Trans. Intell. Syst. Technol.}}
  \bibinfo{volume}{8}, \bibinfo{number}{4}, Article \bibinfo{articleno}{58}
  (\bibinfo{date}{Feb.} \bibinfo{year}{2017}), \bibinfo{numpages}{15}~pages.
\newblock
\showISSN{2157-6904}
\urldef\tempurl%
\url{https://doi.org/10.1145/3011871}
\showDOI{\tempurl}


\bibitem[\protect\citeauthoryear{Yin, Hu, Tang, Daly, Zhou, Ouyang, Chen, Kang,
  Deng, Nobata, et~al\mbox{.}}{Yin et~al\mbox{.}}{2016}]%
        {yin2016ranking}
\bibfield{author}{\bibinfo{person}{Dawei Yin}, \bibinfo{person}{Yuening Hu},
  \bibinfo{person}{Jiliang Tang}, \bibinfo{person}{Tim Daly},
  \bibinfo{person}{Mianwei Zhou}, \bibinfo{person}{Hua Ouyang},
  \bibinfo{person}{Jianhui Chen}, \bibinfo{person}{Changsung Kang},
  \bibinfo{person}{Hongbo Deng}, \bibinfo{person}{Chikashi Nobata},
  {et~al\mbox{.}}} \bibinfo{year}{2016}\natexlab{}.
\newblock \showarticletitle{Ranking relevance in yahoo search}. In
  \bibinfo{booktitle}{\emph{Proceedings of the 22nd ACM SIGKDD International
  Conference on Knowledge Discovery and Data Mining}}.
  \bibinfo{pages}{323--332}.
\newblock


\bibitem[\protect\citeauthoryear{Zankl, Seuschek, Irazoqui, and
  Gulmezoglu}{Zankl et~al\mbox{.}}{2018}]%
        {zankl2018side}
\bibfield{author}{\bibinfo{person}{Andreas Zankl}, \bibinfo{person}{Hermann
  Seuschek}, \bibinfo{person}{Gorka Irazoqui}, {and} \bibinfo{person}{Berk
  Gulmezoglu}.} \bibinfo{year}{2018}\natexlab{}.
\newblock \showarticletitle{Side-Channel Attacks in the Internet of Things:
  Threats and Challenges}.
\newblock In \bibinfo{booktitle}{\emph{Solutions for Cyber-Physical Systems
  Ubiquity}}. \bibinfo{publisher}{IGI Global}, \bibinfo{pages}{325--357}.
\newblock


\bibitem[\protect\citeauthoryear{Zhang, Goh, Win, and Thing}{Zhang
  et~al\mbox{.}}{2016}]%
        {zhang2016image}
\bibfield{author}{\bibinfo{person}{Ying Zhang}, \bibinfo{person}{Jonathan Goh},
  \bibinfo{person}{Lei~Lei Win}, {and} \bibinfo{person}{Vrizlynn~LL Thing}.}
  \bibinfo{year}{2016}\natexlab{}.
\newblock \showarticletitle{Image Region Forgery Detection: A Deep Learning
  Approach.}
\newblock \bibinfo{journal}{\emph{SG-CRC}}  \bibinfo{volume}{2016}
  (\bibinfo{year}{2016}), \bibinfo{pages}{1--11}.
\newblock


\bibitem[\protect\citeauthoryear{Zhou and Standaert}{Zhou and
  Standaert}{2019}]%
        {zhou2019deep}
\bibfield{author}{\bibinfo{person}{Yuanyuan Zhou} {and}
  \bibinfo{person}{Fran{\c{c}}ois-Xavier Standaert}.}
  \bibinfo{year}{2019}\natexlab{}.
\newblock \showarticletitle{Deep learning mitigates but does not annihilate the
  need of aligned traces and a generalized resnet model for side-channel
  attacks}.
\newblock \bibinfo{journal}{\emph{Journal of Cryptographic Engineering}}
  (\bibinfo{year}{2019}), \bibinfo{pages}{1--11}.
\newblock


\end{thebibliography}

\end{document}